\pdfoutput=1
\documentclass[twocolumn]{aastex63}
\usepackage{natbib}
\usepackage{hyperref}
\usepackage{amsmath}
\usepackage{float}
\usepackage{threeparttable}

\newcommand{\Msolar}{$M_{\odot}$}

\newcommand{\simi}{$\sim$}

\newcommand{\kps}{km s$^{-1}$}

\newcommand{\htwo}{H\,{\sc ii}}
\newcommand{\hone}{H\,{\sc i}}

\newcommand{\LxbLy}{$\frac{{\rm{Length}}^{\rm{Fil}}_{\rm{x}}}{{\rm{Length}}^{\rm{Fil}}_{\rm{y}}}$}

\shorttitle{CCC: Formation of HFSs and associated gas kinematics}
\shortauthors{Maity et al.}

\begin{document}
%
\title{Cloud-Cloud Collision: Formation of Hub-Filament Systems and Associated Gas Kinematics \\ Mass-collecting cone: A new signature of Cloud-Cloud Collision} 
\correspondingauthor{A.~K.  Maity}
\email{Email: aruokumarmaity123@gmail.com}
\author[0000-0002-7367-9355]{A.~K. Maity}
\affiliation{Astronomy \& Astrophysics Division, Physical Research Laboratory, Navrangpura, Ahmedabad 380009, India}
\affiliation{Indian Institute of Technology Gandhinagar Palaj, Gandhinagar 382355, India}

\author{T.~Inoue}
\affiliation{Department of Physics, Nagoya University, Furo-cho, Chikusa-ku, Nagoya 464-8601, Japan}

\author[0000-0002-8966-9856]{Y.~Fukui}
\affiliation{Department of Physics, Nagoya University, Furo-cho, Chikusa-ku, Nagoya 464-8601, Japan}

\author[0000-0001-6725-0483]{L.~K. Dewangan}
\affiliation{Astronomy \& Astrophysics Division, Physical Research Laboratory, Navrangpura, Ahmedabad 380009, India}

\author[0000-0003-2062-5692]{H.~Sano}
\affiliation{Faculty of Engineering, Gifu University, 1-1 Yanagido, Gifu 501-1193, Japan}

\author[0000-0002-1865-4729]{R. I. ~Yamada}
\affiliation{Department of Physics, Nagoya University, Furo-cho, Chikusa-ku, Nagoya 464-8601, Japan}

\author[0000-0002-1411-5410]{K.~Tachihara}
\affiliation{Department of Physics, Nagoya University, Furo-cho, Chikusa-ku, Nagoya 464-8601, Japan}

\author[0000-0001-8812-8460]{N.~K.~Bhadari}
\affiliation{Astronomy \& Astrophysics Division, Physical Research Laboratory, Navrangpura, Ahmedabad 380009, India}

\author[0009-0001-2896-1896]{O.~R.~Jadhav}
\affiliation{Astronomy \& Astrophysics Division, Physical Research Laboratory, Navrangpura, Ahmedabad 380009, India}
\affiliation{Indian Institute of Technology Gandhinagar Palaj, Gandhinagar 382355, India}

\begin{abstract}
Massive star-forming regions (MSFRs) are commonly associated with hub-filament systems (HFSs) and sites of cloud-cloud collision (CCC). Recent observational studies of some MSFRs suggest a possible connection between CCC and the formation of HFSs. To understand this connection, we analyzed the magneto-hydrodynamic simulation data from Inoue et al. (2018). This simulation involves the collision of a spherical turbulent molecular cloud with a plane-parallel sea of dense molecular gas at a relative velocity of about 10 {\kps}. Following the collision, the turbulent and non-uniform cloud undergoes shock compression, rapidly developing filamentary structures within the compressed layer. We found that CCC can lead to the formation of HFSs, which is a combined effect of turbulence, shock compression, magnetic field, and gravity. The collision between the cloud components shapes the filaments into a cone and drives inward flows among them. These inward flows merge at the vertex of the cone, rapidly accumulating high-density gas, which can lead to the formation of massive star(s). The cone acts as a mass-collecting machine, involving a non-gravitational early process of filament formation, followed by gravitational gas attraction to finalize the HFS. The gas distribution in the position-velocity ($PV$) and position-position spaces highlights the challenges in detecting two cloud components and confirming their complementary distribution if the colliding clouds have a large size difference. However, such CCC events can be confirmed by the $PV$ diagrams presenting gas flow toward the vertex of the cone, which hosts gravitationally collapsing high-density objects, and by the magnetic field morphology curved toward the direction of the collision. 
\end{abstract}
%
\keywords{
ISM: clouds -- ISM: kinematics and dynamics -- stars: formation -- stars: massive
stars
}
%
\section{Introduction}
\label{sec:intro_JP}
Despite being rare, massive stars ($M$ $\gtrsim$ 8\,{\Msolar}) greatly influence the evolution of their host galaxies due to their immense radiative and mechanical feedback \citep[][and references therein] {motte_2018}. They are also responsible for the chemical enrichment of the interstellar medium \citep{Dunne2003Nature,Du_2021RAA}. However, while the impact of massive stars on galactic evolution and interstellar chemistry is well-documented, their formation mechanisms remain poorly understood. Significant progress has been achieved in the last few decades using advanced telescopes and computing facilities for observational and theoretical/numerical works in this field. However, a complete understanding of massive star formation (MSF) is still elusive \citep[e.g.,][]{zinnecker07, tan14, Krumholz_2015ASSL, motte_2018}, and demands further studies.

Earlier works have well established that MSF requires a high-density region, having a H$_2$ column density, $N$(H$_2$) $\gtrsim$ 10$^{22}$ cm$^{-2}$ \citep{Krumholz_2007,zinnecker07,fukui21}. Such high column density regions are commonly found at the junction of several filaments \citep[e.g.,][]{myers_2009,trevino19,Kumar_2020,Dewangan_2023JApA} or at the interface of colliding molecular clouds \citep[e.g.,][]{torri_2017,Sano_2021PASJ,fukui18b,fukuia_2018,fujita21,maity_W31,Maity_2023MNRAS}. 
In the literature, the former scenario is based on accretion through filaments to their common junctions, known as hubs, and such systems are referred to as hub-filament systems \citep[HFSs;][]{myers_2009}. On small-scales (i.e., $<$ 1 pc), mass accretion through filaments to their hub is primarily driven by gravity. However, on larger-scales (i.e., $>$ 1 pc), both turbulence as inertial flow and gravitational contraction can contribute to the process \citep[e.g.,][]{Zhou_2022MNRAS}. Thus, hubs accumulate substantial amounts of material through the filaments during the early stages of evolution and play a crucial role in initiating the formation of star clusters, including massive stars \citep[][]{schenider_2012, motte_2018, Kumar_2020}.
A comprehensive understanding of the Galactic HFSs can be achieved from the study of \citet{Kumar_2020}. They analyzed around 35,000 Herschel clumps and discovered about 3,700 candidate HFSs. Additionally, their findings indicate that almost all the highest mass stars ($>$ 100 {\Msolar}) form in the hubs.
The latter scenario is based on the collision of molecular clouds, which is known as cloud-cloud collision \citep[CCC;][]{habe92}. 
The CCC rate is proportional to the molecular cloud-cloud velocity dispersion \citep[see Equation 2 in][]{dobbs_2015} in the Galactic disk, which is reported to be around 6 {\kps} \citep[e.g.,][]{Jog_1988ApJ,Stark_2005ApJ,dobbs_2015,Inutsuka_2015A&A}.
In general, CCC sites show the bridge feature and complementary distribution in the position-velocity ($PV$) and position-position ($PP$) space, respectively \citep[e.g.,][]{torri_2011,torri_2015,torri_2017,torri_2021,fukui_2014,fukui_2015,dewangan_2017,nishimura_2017,hayashi_2018,Sano_2021PASJ,sano_2018,fujita21,Maity_2023MNRAS}. The bridge feature is a subtle connection of the colliding cloud components in $PV$ space \citep[e.g.,][]{haworth_2015,torri_2017,Dewangan2017,Priestley2021}. In $PP$ space, the spatial fit of ‘intensity-enhancement’ and ‘intensity-depression’ regions of the cloud components is known as the complementary distribution \citep{fukui21,fujita21}. In the case of a supersonic collision, the effective sound speed increases at the shock-compressed interface of the colliding clouds \citep{habe92,anathpindika_2010,fukui_2014,fukuia_2018}. The increase in effective sound speed provides a higher mass accretion rate of around $10^{-4}$--$10^{-3}$ {\Msolar} yr$^{-1}$ \citep[][]{inoue13}, leading to the formation of massive stars. A detailed theoretical and observational overview of CCC can be found in \citet{fukui21}, which also includes the list of more than 50 observationally detected CCC sites and their corresponding physical parameters. 
However, in addition to the above-mentioned mechanisms, physical processes such as the gravitational collapse of molecular clouds, filament fragmentation, and edge-dominated collapse can also lead to high-density regions suitable for MSF \citep[e.g.,][and references therein]{Schneider_2015A&A,Beuther_2015A&A,Bhadari_2020,Bhadari_2022ApJ}.

Recent observational work on Galactic massive star-forming regions (MSFRs), specifically SDC13 \citep{Peretto_2014,Wang_2022ApJ}, W31 complex \citep{maity_W31}, W33 complex \citep{Zhou_2023MNRAS}, and AFGL 5180 \& 6366S \citep{Maity_2023MNRAS}, reveals simultaneous signatures of CCC and the presence of HFSs. Similar results have also been reported for extragalactic MSFRs, such as N159E-Papillon \citep{fukui19ex}, N159W-South \citep{tokuda19ex}, and 30 Dor \citep{Wong_2022ApJ} in the Large Magellanic Cloud (LMC). Although the literature provides limited instances of MSFRs exhibiting concurrent signatures of CCC and the presence of HFSs, this raises two important questions regarding MSF: Can CCC lead to the formation of HFS? If so, how do filaments precisely converge toward a common junction after the collision, resulting in the observed morphology of the HFSs? In this context, a connection between CCC and the formation of a network of filaments was reported in smoothed particle hydrodynamics simulation by \citet{balfour15,Balfour_2017MNRAS}. \citet{balfour15} showed that lower collision velocities of uniform density clouds with subsonic turbulence form radially converging filaments (referred to as `hub-and-spoke' system), while relatively higher collision velocities lead to a complex network of filaments resembling a spider’s web. The distinction between the `hub-and-spoke' system and the spider's web structure becomes blurred for the collisions of molecular clouds with internal substructure \citep{Balfour_2017MNRAS}.  
Their works, however, did not delve deeply into the formation of HFSs. In addition, the absence of magnetic fields in their research can be recognized as a major limitation. The formation of filaments in the CCC scenario was extensively explored using magneto-hydrodynamic (MHD) simulations by \citet{inoue13} and \citet{inoue18}. They demonstrated that shock compression of a turbulent and inhomogeneous cloud leads to the rapid development of filaments. However, it is yet to be explored whether these filaments converge over time to form a HFS and what potential factors influence this process.

Hence, this paper aims to analyze the MHD simulation data generated by \citet{inoue18} to unravel the formation mechanism of HFSs through CCC. Specifically, it examines the roles of turbulence, gravity, and magnetic field in the entire process. Additionally, this simulation data provide valuable insights into the intricate gas kinematics involved in CCC, aiding in understanding the observational challenges in detecting CCC sites and proposing new signatures to identify them. This paper is organized as follows: Section~\ref{sec:model_JP} describes the details of the simulation data utilized in this work. Section~\ref{sec:results_JP} presents the results of our analysis, which are further discussed in Section~\ref{sec:dis_JP}. Section~\ref{sec:sum_JP} summarizes the major findings of this work.
\section{The model description}
\label{sec:model_JP}
The simulation of CCC, conducted by \citet{inoue18}, incorporated ideal MHD with self-gravity and utilized adaptive mesh refinement and sink particle techniques \citep[developed by][]{Matsumoto_2007PASJ,Matsumoto_2015ApJ}. Poisson’s equation was solved for gravitational potential energy, and the negative gradient of the gravitational potential provides the gravitational acceleration. Gravity was present in the system from the beginning of the simulation. The molecular gas was approximated to be isothermal, corresponding to the sound speed ($c_\mathrm{s}$) of about 0.3 {\kps}. The cubical numerical domain has each side of 6 pc, with its center at the origin (x = 0, y = 0, z = 0). As an initial setup, a spherical molecular cloud with a radius ($R$) of 1.5 pc was placed at (0, 0, 0.5 pc) with a z-component of velocity ($v_\mathrm{z}$) of about $-$5 {\kps}, as shown in Figure~\ref{initial_density_dist}a. The initial density ($n$) distribution of the spherical cloud was a Gaussian with an amplitude of $10^{3}$ cm$^{-3}$ and a full width at half maximum (FWHM) of about 2.5 pc.

The mean molecular mass of gas particles was set to be 2.4 times the proton mass. Hence, the total mass of the spherical cloud was about 477\,{\Msolar}.
Turbulence is a common feature in molecular clouds \citep{Larson_1981MNRAS}, potentially arising from larger-scale flow collisions, Galactic shear, supernova explosions, expanding {\htwo} regions, protostellar outflows, and stellar winds \citep[e.g.,][]{Semadeni_2003ASPC,Hennebelle_2012A&ARv,Kim_2015ApJ,Padoan_2016ApJ,Orkisz_2017A&A}. 
Therefore, in addition to the bulk motion of the spherical cloud, a turbulent velocity field ($v_t$) with a FWHM, $\Delta v_t \sim1$ {\kps} was incorporated.
As the second cloud component, a sea of dense gas with $n$ = 10$^3$ cm$^{-3}$ was set in the region z $ \lesssim -1.5$ pc (see Figure~\ref{initial_density_dist}a) with $v_\mathrm{z}$ of about 5 {\kps}. The remaining space was filled with relatively less dense gas with $n$ = 10$^2$ cm$^{-3}$. 
The histogram of CCC events with the relative velocity between the colliding clouds \citep[i.e., the collision velocity; see Figure 9b in][]{fukui21} reveals the mean collision velocity to be about 5 {\kps}. However, the cloud components observed in all the CCC sites are in collisional interaction, which decreases the velocity separation between the cloud components \citep[][]{takahira_2014}. In addition, the projection effect can reduce the observed velocity separation between the cloud components. Hence, the actual collision velocity will be larger, so a collision velocity of about 10 {\kps} is used in this work. This CCC model resembles a scenario in which a larger cloud collides with a smaller one or a cloud undergoes compression by a plane-parallel shock wave{\footnote{The interaction of the molecular cloud with an expanding bubble can act as a source of plane-parallel shock \citep{Inutsuka_2015A&A}.}}. The initial magnetic field, $\vec{B} = [0, 20\,\mu{\mathrm{G}}, 0]$, was oriented perpendicular to the direction of the collision. The input density and z-component of velocity distributions along the z-axis are shown in Figures~\ref{initial_density_dist}b and ~\ref{initial_density_dist}c, respectively. These distributions can be mathematically described as follows:
\begin{equation}
\label{eq1}
n (0, 0, z) = \begin{cases}
550 - 450 \,\tanh (\frac{z + 1.5}{0.1}) &\text{if }  z<-1\, \mathrm{and} \,z >2, \\
1000 \exp (-\frac{(z-0.5)^2}{2\sigma^2_z}) &\text{if }  -1\leq z \leq 2 ,
\end{cases}
\end{equation}
and 
\begin{equation}
 v_z (0, 0, z) = -5\,\tanh (\frac{z + 1.5}{0.1}) + v_t (0, 0, z), 
\label{eq2} 
\end{equation}
where, $\sigma_z = \frac{2.5\,\mathrm{pc}}{\sqrt(8\ln2)}$ is the dispersion of the spherical cloud. Equations~\ref{eq1} and \ref{eq2} are in the units of cm$^{-3}$ and km s$^{-1}$, respectively, with z measured in pc in both cases. The numerical results of the simulation were recorded from time, $t$ = 0 to 0.7 Myr at an interval of 0.1 Myr. The data utilized in this work have a resolution of about $\frac{6}{512}$ pc = 0.0117 pc {\simi}2400 AU, which is sufficient for the study of HFSs and the associated gas kinematics.
\section{Results}
\label{sec:results_JP}
\subsection{The physical environment of the shock-compressed layer and distribution of the sink particles}
\label{sec:shock_env_JP}
The evolution of the system from $t$ = 0.1 to 0.7 Myr is presented in Figure~\ref{NH2} using the column density (i.e., $N$(H$_2$)) maps in the x-z, x-y, and y-z planes. The simulation introduces sink particles to follow star formation for regions above the density threshold of about 10$^{6}$ cm$^{-3}$, following the criteria detailed by \citet{inoue18}, which include a local potential minimum, negative velocity divergence, negative eigenvalues of the velocity gradient tensor, and negative total energy within the sink radius. The positions of the sink particles created in this simulation are displayed on the column density maps using the plus ($+$) symbols.
In the x-z and y-z planes, the column density maps show the development of the shock-compressed layer. As both planes are parallel to the direction of the collision, they present similar gas distribution. Hereafter, we mostly used the y-z plane to show the gas distribution from a perpendicular direction to the collision axis (i.e., the z-axis). The x-y projected maps reveal structural components (i.e., filaments and cores) of the compressed layer. Here, the term `core' refers to gaseous condensations that do not meet the selection criteria for the sink particles. The structural components and the sink particles are mostly confined within the circular regions, as shown with the gray dashed circles in Figure~\ref{NH2}. These circular areas are selected for further examination (see Sections~\ref{sec:fil_detect_JP}, \ref{sec:ang_BVG_JP} and Figures~\ref{NH2_Filaments+cores}, \ref{Filaments_with_BVG}). The pair of white lines overlaid on the x-y projected column density map at $t$ = 0.3 Myr along the x- and y-axes are employed to extract the $PV$ diagrams (see Section~\ref{sec:CCC_sign_JP} and Figure~\ref{PV_withlies}).
The histograms of the gas density for each pixel within z: $[-2, 2]$ pc and $n$: [5$\times$10$^1$, 10$^7$] cm$^{-3}$ are displayed in Figure~\ref{NH2_hist}, with $t$ = 0 Myr showing the input distribution. A low-density gas distribution below the lowest input density (i.e., 10$^2$ cm$^{-3}$, marked with blue dotted lines in Figure~\ref{NH2_hist}) is observed for $t$ = 0.1 to 0.7 Myr. Turbulence is responsible for the creation of these low-density regions \citep{Fukui_2021PASJ}. 
Several numerical studies have shown that supersonic turbulence in an isothermal uniform density molecular cloud leads to a lognormal density distribution \citep[e.g.,][and references therein]{Federrath_2010A&A,Auddy_2018MNRAS}, which has densities both above and below the initial density. Initially, the cloud was geometrically symmetrical, moving with uniform velocity and in a uniform magnetic field. Hence, gravity and the magnetic field will not be responsible for the production of low-density regions. Therefore, turbulence is the only factor responsible for the production of those low-density regions.
At the same time, the collision of cloud components produces high-density regions, characterized by a power-law tail above {\simi}$2 \times 10^4$ cm$^{-3}$, expressed as $N_{\mathrm{pix}} \propto n^{-p}$. Here, $N_{\mathrm{pix}}$ presents the pixel number at a given density $n$, and $p$ is the power-law index. It is evident in Figure~\ref{NH2_hist} that the value of $p$ decreases from 7.54 at $t$ = 0.2 Myr to 1.99 at $t$ = 0.7 Myr.  Consequently, the high-density tail gradually flattens over time.

The temporal change in the cone-shaped{\footnote{In real life, molecular clouds are not perfect Gaussians. As long as the central region of a molecular cloud is the densest and the cloud collides with a larger cloud component, a conical structure can be expected. The resulting cones, though, may be imperfect or distorted based on the initial structure of the molecular cloud.}} compressed layer is displayed in the y-z projected column density maps for an integration range of x: $[-1,1]$ pc in Figure~\ref{cones_fig}. The compressed layer at each time exhibits symmetry about the collision axis and its angle ($\theta$) with the z-axis decreases from $\theta$ $\sim$84$^{\circ}$ at $t = 0.2$ Myr to $\theta$ $\sim$52$^{\circ}$ at $t = 0.7$ Myr due to the collisional impact (see the angles in Figure~\ref{cones_fig}).
All the cones presented in Figure~\ref{cones_fig} are divided into thin conical shells having a width of 3 pixels. These thin shells (in 3D) are utilized to calculate the mass-weighted average density, z-component of velocity, and the gravitational acceleration.
The results of this analysis are shown in Figure~\ref{nvz_dist}, which demonstrate the rapid development of high-density and intermediate-velocity regions. The width and density of the compressed layer increase over time,  which leads to an increase in the gravitational acceleration with time. Finally, the compressed layer grows to a thickness of about 1 pc at $t$ = 0.7 Myr.
The detected z-component of velocity magnitudes above the highest input value, i.e., \(|v_\mathrm{z}|\) $ \gtrsim 5$ {\kps} result from gravitational acceleration.
The high-density gas distribution for the conical geometries of the compressed layers along the z-direction from $t$ = 0.2 to 0.7 Myr is shown in Figure~\ref{hd_dist}. At $t$ = 0.3 and 0.4 Myr, the maximum gas density reaches more than 10$^5$ and 10$^6$ cm$^{-3}$, respectively, which can also be inferred from the $n_{\mathrm{max}}$ values mentioned in Figure~\ref{NH2_hist} and the detection of sink particles after $t$ = 0.4 Myr. A smooth development of the high-density region with $n > 10^4$ cm$^{-3}$ along the z-direction can be seen. However, the distribution of gas with the density, $n > 10^5$ cm$^{-3}$ along the z-direction shows large variations due to the formation of structural components, which are described later.
Table~\ref{tab1} presents the total mass of the gas within the compressed layer with the density exceeding 10$^3$ cm$^{-3}$. We observe an increase in the fraction of dense gas over time. Specifically, the fraction of dense gas with $n > 10^5$ cm$^{-3}$ to the total mass of gas with $n > 10^4$ cm$^{-3}$ rises from about 0.1\% at $t$ = 0.3 Myr to approximately 18\% at $t$ = 0.7 Myr. This ratio will increase further if the contributions of the sink particles are added to the calculation. 

Figure~\ref{04_2dhist} presents the distribution of various physical parameters throughout the compressed layer at $t$ = 0.4 Myr. The parameters include velocity magnitude ($|\vec{V}|$), the angle between the velocity vector and the z-axis, density, dynamic pressure, and the magnitude of the magnetic field ($|\vec{B}|$). As a result of the collision, a high-density compressed layer emerges, characterized by a small velocity ($|\vec{V}| \lesssim 2$ {\kps}), having a wide range of angles with respect to the axis of collision. This transformation is attributed to shock dissipation and momentum exchange between the colliding components \citep{Fukui_2021PASJ}.
However, a significant amount of gas is moving at large angles relative to the collision axis (see the black contours), which is also consistent with Figure~\ref{cones_fig}. The black contours represent 30\%, 60\%, and 90\% of the total number of pixels, excluding contributions from the initial gas distribution and gravitationally accelerated gas. Additionally, the compressed layer exhibits a broad range of dynamic pressure and magnetic fields. The colliding motion pinches the magnetic field lines, amplifying the field strength to around 0.3--1 mG. 
%
\begin{table}
\centering
\caption{Total mass of the dense gas. Density, $n$ is in cm$^{-3}$.}
\label{tab1}
\begin{tabular}{ccccc}
\hline  \vspace{0.1cm}
    &          & Mass ({\Msolar})  &   \\
\cline{2-5}
$t$ (Myr) & $n> 10^{3}$ & $n> 10^{4}$ & $n>10^{5}$ & $n>10^{6}$\\
\hline 
0.2 &  337 &  262 &{---}& {---} \\
0.3 &  937 &  668 &   5 & {---} \\
0.4 & 1335 & 1037 &  40 &     1 \\
0.5 & 1743 & 1196 & 117 &     6 \\
0.6 & 1959 & 1302 & 178 &    13 \\
0.7 & 2149 & 1439 & 260 &    27 \\

\hline
\end{tabular}
\end{table}
%
\subsection{Detection of the filaments and cores}
\label{sec:fil_detect_JP}
As the clouds collide along the z-axis, the shock-compressed layer develops primarily in the x-y plane. The dense gas distributions, similar to those in Figure~\ref{hd_dist} but along the x- and y-axes (corresponding plots are not included here), show spike-like features for $n > 10^5$ and $> 10^6$ cm$^{-3}$. This suggests that the structural
components forming in the compressed layer are distinctly visible in the x-y plane. Hence, the x-y projected column density maps were used for detecting filaments and cores embedded in the compressed layer from $t$ = 0.2 to 0.7 Myr. Firstly, the column density maps were smoothed to a 3-pixel resolution using  {\it convolve}\footnote[1]{The {\it convolve} task is available with {\it getsf} \citep{getsf_2022}.}. Then, we applied the {\it getsf} algorithm \citep{getsf_2022} to the smoothed column density maps to extract embedded filaments and cores.
The {\it getsf} breaks down an image into its structural components, distinguishing sources and filaments and isolating them from each other and their backgrounds. Additional information about this algorithm is available in \citet{getsf_2022}. Essential inputs for the algorithm include the maximum size of structural elements (i.e., filaments and cores) and the angular resolution of the image. In this case, the structural components are extracted from the column density maps, with maximum filament and core sizes of 8 and 6 pixels, respectively.

The filament skeletons and the cores detected on global-scale in the {\it getsf} utility are highlighted over the column density maps in Figure~\ref{NH2_Filaments+cores}. Multiple filament skeletons are identified from $t$ = 0.2 to 0.7 Myr within our target regions (see the gray dashed circles in Figure~\ref{NH2}). Notably, no cores are detected at $t$ = 0.2 Myr or earlier. The high-density skeletons with $N(\mathrm{H}_2)^\mathrm{crest}_\mathrm{med} \geq 10^{21.7}$ cm$^{-2}$ are marked in green, while the low-density skeletons are in white. The column density threshold for the high-density skeletons is set based on the studies of \citet{Planck_2016_feb} and \citet{Cox_2016AA}. Here, $N(\mathrm{H}_2)^\mathrm{crest}_\mathrm{med}$\footnote[2]{The {\it getsf}-catalogs provide $N(\mathrm{H}_2)^\mathrm{crest}_\mathrm{med}$, length, mass, and position angle of the filaments, as well as the size and mass of the cores. We adjusted the mass values using a mean molecular mass of 2.4, which was initially set at 2.8.} is the median column density value along the filament skeleton \citep{getsf_2022}. The cores and the sink particles are primarily found to be projected either over the spine of the high-density filaments or at their common junctions (see Figure~\ref{NH2} for the sink particles). Such distributions are detected toward several star-forming regions, including IC 5146 Dark Streamer \citep{Dewangan2023ApJ} and RCW 117 \citep{Seshadri_2024MNRAS}. Hence, in particular these high-density filaments are involved in the star formation activity. The details of the high-density filaments from $t$ = 0.2 to 0.7 Myr are summarized in Table~\ref{tab2}. Their number and combined length increase until $t$ = 0.5 Myr, after which a significant decrease occurs. This trend is also evident in the total mass of these filaments. The projection effect will result in a maximum increase of about 27\% in the calculated length for $\theta $ $\sim$52$^{\circ}$ at $t = 0.7$ Myr. We introduced a quantity, denoted as {\LxbLy}, to indicate the orientation of the high-density filaments relative to the initial magnetic field. Here, {\LxbLy} represents the ratio of the total projected length of the high-density filaments along the x- and y-axis, i.e., perpendicular and parallel to the initial magnetic field, respectively. Therefore, if {\LxbLy} $\ll$ 1, filaments are largely oriented along the initial magnetic field, and for {\LxbLy} $\gg$ 1, filaments are mainly perpendicular to the initial magnetic field. At $t$ = 0.3, 0.6, and 0.7 Myr, {\LxbLy} $>$ 1.5 indicates that the dense filaments are more aligned perpendicular to the initial magnetic field. The possible reasons and implications of the variation of {\LxbLy} are discussed in detail in Section~\ref{sec_shock_env_JP_dis}. The collective properties of the {\it getsf}-identified cores and the sink particles from $t$ = 0.2 to 0.7 Myr are presented in Table~\ref{tab3}. The total mass of the cores and the sink particles steadily increases as time progresses. The spectrum of mass for both the cores and the sink particles broadens over time. At each timestep, the mass of the heaviest core, depicted by a blue circle with a white edge in Figure~\ref{NH2_Filaments+cores}, increases, ultimately yielding a massive core of about 15\,{\Msolar} at $t$ = 0.7 Myr. The mass of the heaviest sink particle also increases over time and reaches about 40\,{\Msolar} at $t$ = 0.7 Myr (see the white plus symbol in Figure~\ref{NH2}). Additionally, at $t$ = 0.7 Myr, the heaviest sink particle is accompanied by a massive companion of about 11\,{\Msolar}, located at the junction of the filaments (see the cyan plus symbol in Figure~\ref{NH2}).
%
\begin{table}
\centering
\caption{Physical parameters for the high-density filaments with $N$(H$_2$)$^{\mathrm{crest}}_{\mathrm{med}} \geq 10^{21.7}$ cm$^{-2}$.}
\label{tab2}
\begin{tabular}{ccccc}
\hline  
 $t$ (Myr) & $N^{\rm{Fil}}$& Length$^{\rm{Fil}}_{\rm{tot}}$ (pc) & Mass$^{\rm{Fil}}_{\rm{tot}}$  ({\Msolar})& {\large{\LxbLy}}\\
\hline 
0.2 &     2 &  0.62 &   51 &   0.80 \\
0.3 &     8 &  4.74 &  466 &   1.51 \\
0.4 &    19 & 10.35 &  864 &   1.03 \\
0.5 &    31 & 13.15 & 1189 &   1.17 \\
0.6 &    24 &  8.81 &  780 &   1.54 \\
0.7 &    23 &  8.50 &  983 &   1.60 \\

\hline
\end{tabular}
\end{table}
\begin{table*}
\centering
\caption{Details of the {\it getsf}-identified cores and the sink particles.}
\label{tab3}
\begin{tabular}{cccccccccc}
\hline
$t$ & $N^{\rm{core}}$& $M^{\mathrm{core}}_{\rm{min}}$ & $M^{\mathrm{core}}_{\rm{max}}$  & $M^{\mathrm{core}}_{\rm{total}}$ & $N^{\rm{sink}}$& $M^{\mathrm{sink}}_{\rm{min}}$ & $M^{\mathrm{sink}}_{\rm{max}}$  & $M^{\mathrm{sink}}_{\rm{total}}$ & $M^{\mathrm{core}}_{\rm{total}} + M^{\mathrm{sink}}_{\rm{total}}$\\
(Myr) &  &  ({\Msolar}) &  ({\Msolar}) & ({\Msolar}) & & ({\Msolar})& ({\Msolar})& ({\Msolar}) & ({\Msolar})\\
\hline
0.2 &  {---} &     {---} &     {---} &  {---} &     {---}&  {---}& {---}& {---}&  {---}\\
0.3 &      3 &      0.71 &      1.06 &   2.58 &     {---}&  {---}& {---}& {---}&   2.58\\
0.4 &     19 &      0.24 &      3.88 &  17.05 &     {---}&  {---}& {---}& {---}&  17.05\\
0.5 &     46 &      0.21 &      7.72 &  80.17 &        3 &   0.16&  5.04&  7.60&  87.77\\
0.6 &     50 &      0.36 &      9.81 & 104.74 &        9 &   0.28& 20.25& 31.89& 136.63\\
0.7 &     50 &      0.19 &     15.39 & 123.78 &       25 &   0.04& 40.25& 101.5& 225.28\\

\hline
\end{tabular}
\end{table*}
%
\subsection{The relative orientation of the filament, velocity, magnetic field and gravitational field vectors}
\label{sec:ang_BVG_JP}
To comprehend the formation mechanism of HFS, it is essential to understand the influence of turbulence, magnetic field, and gravity on the flow of materials through the filaments. Hence, the x-y projected mass-weighted average velocity ($\vec{V}$), magnetic field ($\vec{B}$), and gravitational field vectors ($\vec{g}$) toward the filaments are depicted in Figure~\ref{Filaments_with_BVG} from $t$ = 0.3 to 0.7 Myr. These vectors are overlaid on the column density maps for the circular regions highlighted in Figure~\ref{NH2}. The vectors precisely at the skeleton (i.e., the spine) of the high-density filaments are shown in black. In contrast, those in the immediate surrounding pixels of the skeletons, selected based on column density thresholds on the {\it getsf}-identified filament images, are shown in white. To obtain a clear picture of their orientations, $\vec{B}$, $\vec{V}$, and $\vec{g}$ are plotted at intervals of a few pixels in both the x- and y-directions in Figure~\ref{Filaments_with_BVG}. A parallel velocity vector along the filament skeleton (see the black arrows) indicates material flow through the filaments, while perpendicular white arrows depict materials from the surrounding area being fed into the filaments \citep[e.g.,][]{Gomez_2018,Wang_2022ApJ}. However, a perpendicular velocity field at the spine of the filaments reflects their overall bulk motion. Both types of configurations can be observed through visual examination of the velocity vectors in Figure~\ref{Filaments_with_BVG}. We observe a gradual increase in the strength of the $\vec{B}$ and $\vec{g}$. The magnetic field vectors, which typically make large angles with the filaments, reflect their significant role in filament formation \citep[e.g.,][]{inoue13,Pillai_2015ApJ,inoue18,Wang_2022ApJ}. The small angles between $\vec{V}$ and $\vec{g}$ indicate the considerable influence of gravity on the gas kinematics. 

In this context, it is essential to perform a quantitative measurement of the relative orientations (i.e., the difference in the position angle, $\Delta$PA) among the filaments, $\vec{B}$, $\vec{V}$, and $\vec{g}$. Hence, Figure \ref{Delta_PA_dist} presents histograms of the relative orientation among the filaments, $\vec{B}$, $\vec{V}$, and $\vec{g}$ at $t$ = 0.7 Myr. The supplementary plots for $t$ = 0.3 to 0.6 Myr can be found in the Appendix, in Figure~\ref{Delta_PA_dist_app}. All pixels from the high-density skeletons and their surroundings (as mentioned for Figure~\ref{Filaments_with_BVG}) are used in Figures~\ref{Delta_PA_dist} and \ref{Delta_PA_dist_app}.
The large $\Delta$PA between the filament skeletons and $\vec{B}$ supports the idea that filaments easily grow perpendicular to the magnetic field \citep[e.g.,][]{inoue13,Cox_2016AA,inoue18}. However, the smaller $\Delta$PA values arise from the dragging of the magnetic field lines, resulting in more parallel components due to the material flowing through the filaments \citep[e.g.,][]{Gomez_2018}. Thus, the $\Delta$PA between $\vec{B}$ and the filament skeletons varies over a wide range at $t$ = 0.7 Myr. At the same time (i.e., $t$ = 0.7 Myr), similar wide angular ranges for $\Delta$PA between $\vec{V}$ and the filament skeleton are observed due to two factors: first, gas flow along the spine of the filament, resulting in a parallel relative configuration; and second, the bulk motion of the filaments. The $\Delta$PA between $\vec{g}$ and the filament skeleton shows a clear bimodal distribution at $t$ = 0.7 Myr, contrasting with their distribution at $t$ = 0.3 Myr.
The bimodal distribution of $\Delta$PA between the filament skeleton and $\vec{g}$  arises because, toward the central region, materials flow through the filaments toward the hub, giving rise to the parallel component. However, filaments in the outer regions flow as a whole toward the hub, but their orientations are perpendicular to the gravitational field vectors.
In Figure~\ref{Delta_PA_dist_app}, it is evident that $\vec{B}$ forms large angles with $\vec{V}$ for the high-density filaments but appears randomly oriented for the surrounding gas. However, with time, $\vec{V}$ and $\vec{B}$ increasingly align, both for the high-density filaments and their surroundings. This suggests the role of the magnetic field in guiding gas motions in the surroundings and in the bending of magnetic fields due to gas flow through the filaments. 
Similar findings were reported in studies by \citet{Pillai_2015ApJ,Juarez_2017ApJ,Klassen_2017MNRAS,Liu_2018ApJ,Gomez_2018,Koch_2022ApJ}.
The random distribution of $\vec{B}$ with respect to $\vec{g}$ indicates that the magnetic field does not have a significant role compared to gravity.
At $t$ = 0.3 Myr, $\vec{V}$ forms large angles with respect to $\vec{g}$ for the high-density filament skeletons and is nearly randomly distributed with $\vec{g}$ for the surrounding regions. This suggests that gravity does not significantly influence gas kinematics in the initial stages. However, shortly after that, $\vec{V}$ aligns with $\vec{g}$ for both the skeleton and the surrounding regions (see Figure~\ref{Delta_PA_dist} and \ref{Delta_PA_dist_app}), implying that gravity becomes dominant in governing gas kinematics during later stages of evolution, as observed toward MSFR SDC13 \citep{Wang_2022ApJ}.

The main objective of this study is to investigate the influence of magnetic fields and gravity on gas flow within filaments. To achieve this, we selected high-density filament skeletons and their surrounding regions at each timestep to analyze $\Delta$PAs among the filaments, $\vec{B}$, $\vec{V}$, and $\vec{g}$. Core formation activities may alter the $\Delta$PAs on scales smaller than 0.1 pc, but this is not expected to impact the filament scale \citep[e.g.,][]{Wang_2022ApJ}. In their observational studies, \citet{Wang_2022ApJ} and \citet{Liu_2023ApJ} emphasized that the gravitational field may have a very different orientation than the magnetic field and velocity in the diffuse regions far from the filament skeletons, suggesting that the magnetic field, rather than gravity, influences the low-density areas.

\subsection{The gas kinematics and the spatial distribution of different velocity components}
\label{sec:CCC_sign_JP}
As outlined in Section~\ref{sec:intro_JP}, molecular clouds commonly collide with each other and several theoretical and numerical studies estimated the rate of CCC to be once in every 100 years for our Galaxy \citep[e.g.,][]{tasker_2009, dobbs_2015}.
Despite the large theoretical expectations, the observational record of CCC is limited, with fewer than 100 instances documented in \citet{fukui21}. Therefore, this analysis aims to understand the observational difficulties in the detection of CCC. Using the simulation data, we generated a position(x)-position(y)-velocity($v_\mathrm{z}$) data cube for $t$ = 0 to 0.7 Myr. The velocity interval between the channels is set to be 0.5 {\kps}, which is consistent with several Galactic CO Surveys as listed in \citet{Park_2023_apj}. These data cubes are analogous to the observed molecular line data used for tracing molecular gas kinematics. The data cubes are utilized to extract the $PV$ diagrams along the x- and y-axes for the regions shown in Figure~\ref{NH2}i. The $PV$ diagrams at $t$ = 0, 0.2, 0.4, 0.5, and 0.7 Myr are shown in Figure~\ref{PV_withlies}. The integration range in the y-direction, $\Delta L_\mathrm{y}$ = 0.4 pc, and in the x-direction, $\Delta L_\mathrm{x}$ = 0.4 pc while extracting $PV$ diagrams along the x- and y-directions, respectively.

At $t$ = 0 Myr, i.e., before the collision, two cloud components are distinctly visible in the $PV$ diagrams (see Figure~\ref{PV_withlies}). However, the growth of the compressed layer can be observed at $t$ = 0.2 Myr, marked by a gray shaded region for $v_\mathrm{z}$: [$-$2.25, 3.75] {\kps} (see the panel on the right). The compressed layer lies between the blue- and red-shifted cloud components, shaded in blue and red for $v_\mathrm{z}$: [$-$6.75,$-$3.75] {\kps} and [4.25, 5.75] {\kps}, respectively. At $t$ = 0.5 Myr, one component of the colliding cloud is entirely embraced within the compressed layer. The red-shifted component and the compressed layer are distinguished by the red and gray shaded regions for $v_\mathrm{z}$: [5.25, 7.25] and [$-$3.25, 4.25] {\kps}, respectively. To observe the distribution of the cloud components in the $PP$ space at $t$ = 0.2 and 0.5 Myr, integrated density maps are generated for the previously defined velocity ranges. The integrated density maps are analogous to the integrated intensity (i.e., moment-0) map of the molecular line data analysis. The integrated density map at $t$ = 0.2 Myr is shown in Figure~\ref{comp_dist}a. The image corresponds to the red-shifted cloud component, while the blue-shifted component and the compressed layer are presented with blue and black contours, respectively. The complementary spatial distribution between the `intensity-depression’ in the integrated density map for the red-shifted component and the `intensity-enhancement’ in integrated density contours for the blue-shifted component is very apparent at $t$ = 0.2 Myr.
At the same location, the intermediate velocity compressed layer is detected in black contours, which is a clear result of CCC. However, at $t$ = 0.5 Myr, when one of the cloud components is completely transformed into the compressed layer, a complementary distribution between the red-shifted cloud component and the compressed layer is observed in Figure~\ref{comp_dist}b. Based on these results, we have discussed the observational challenges in the detection of CCC in Section~\ref{sec:CCC_detect_JP}.

\section{Discussion}
\label{sec:dis_JP}
To address the questions introduced in Section~\ref{sec:intro_JP}, we explored numerical simulation data of a head-on collision incorporating ideal MHD and self-gravity. This collision creates a high-density region at the interface of the colliding clouds, characterized by an intermediate velocity distribution.
Along with the sink particles, several filaments and cores are detected in the high-density interface. The relative orientations of the filament skeletons, $\vec{B}$, $\vec{V}$, and $\vec{g}$ toward the filaments indicate the gas motion and its influencing factors. The $PV$ diagrams demonstrate how the signatures of CCC change over time. All the results are thoroughly discussed in this section. In addition, new observational signatures of CCC at $\theta_\mathrm{col} = 90^{\circ}$ are proposed in Section~\ref{sec:CCC_newsign_JP}.



%
\subsection{The impact of the collision on the molecular cloud components and the formation of filaments}
\label{sec_shock_env_JP_dis}
A supersonic collision between molecular clouds creates a shock-compressed layer at their interface \citep[e.g.,][]{furukawa_2009, Dewangan2017, sano_2018, Enokiya2021, Nishimura2021}. The final density of the interface depends primarily on the initial density of the cloud components and their collisional velocities \citep[see Equation 1 in][]{inoue13}. 
A positive correlation between the collision velocity and the peak column density is presented in \citet{fukui21} for the observational sites of CCC (see Figure 10 in their work).
The width and density of the compressed layer, characterized by an intermediate velocity relative to the colliding cloud components \citep[e.g.,][]{takahira_2014,Baug_2016ApJ,Dewangan_2018ApJ}, increase over time due to shock waves propagating through the cloud components, as discussed in detail by \citet{inoue13}. This increase leads to a rise in the total mass of the high-density gas (see Table~\ref{tab1}) and explains the flattening nature of the high-density tail observed in the histograms of the gas density, as indicated by the red lines in Figure~\ref{NH2_hist}. 
The rapid decrease in the slope (i.e., the flattening nature) of the high-density tail suggests the swift development of high-density pixels in the initial stages after the collision. However, the rate of growth of the high-density region slows down in the later stages due to shock dissipation \citep[e.g.,][]{inoue18, Fukui_2021PASJ}. This phase was described as the `dissipation stage' in the study by \citet{Navarrete_2024arXiv}. They demonstrated the presence of higher Mach numbers at the beginning of the collision. However, the number of shocked cells and their Mach numbers decrease in the later stages. Hence, the production of high-density regions slows down over time in the case of CCC. 
The compressed layer is conducive to the development of structural components, which are favorable environments for star formation \citep[e.g.,][]{Shima_2018PASJ, Fukui_2021PASJ}. 
The initial density distributions, the collision velocity, and the presence of a magnetic field can have a huge impact on the core and filament formation and their properties \citep[e.g.,][]{inoue13,takahira_2014,Arreaga-Garcia2014IJAA,balfour15,Balfour_2017MNRAS,inoue18,Sakre_2021PASJ,Sakre_2023MNRAS}.
When the colliding clouds have pre-collision substructure (i.e., inhomogeneous density distribution), it causes an inhomogeneous shock-compressed layer \citep{inoue13,Balfour_2017MNRAS,inoue18}. In that case, higher collision velocities create more massive cores/sink particles.
\citet{balfour15,Balfour_2017MNRAS} studied particularly low-velocity collisions without considering magnetic fields. The magnetic field strength in our work reaches the order of mG in the plane of the compressed layer, as shown in Figure~\ref{04_2dhist}. Therefore, the role of the magnetic field cannot be ignored in the evolution of the system. After \citet{inoue13} and \citet{inoue18}, this work presents the important roles of magnetic field and high-velocity collision in the formation of filaments. The initial supersonic turbulence (with Mach number $>$ 1) within the spherical cloud creates pre-collision substructure (see Figures~\ref{NH2}a, \ref{NH2}c, and \ref{cones_fig}b) and opposes centralized collapse \citep{Arzoumanian_2011A&A,Federrath_2016MNRAS,Shima_2018PASJ}. Later, the shock compression of the inhomogeneous density structures from the supersonic collision plays a crucial role during/after the collision in the filament formation at the CCC sites. Interestingly, \cite{inoue18} found that comparable filament patterns were formed even in the absence of self-gravity, proving that gravity does not play a role in filament formation, especially if the initial clouds are highly turbulent (see Figure 2 in their paper).

Earlier works \citep[e.g.,][]{inoue13,inoue18} demonstrated that shock compression is efficient when it is parallel to the magnetic field, which preferentially orients the filaments perpendicular to the initial magnetic field \citep[see Figure 3 in ][]{inoue18}. \citet{Abe_2021ApJ} discussed the filament formation mechanisms associated with shock-compressed layers of the molecular clouds and classified the mechanism proposed by \citet{inoue13} as the type-O mechanism \citep[see also][for a review]{Pineda_2023ASPC}. The turbulence-driven inhomogeneous density structures can be randomly oriented with respect to the magnetic field, and the oblique shock effect \citep[for more details see][]{fukui21} will attempt to compress all of them. This results in randomly oriented filaments with respect to the magnetic field. However, as the gas flows along the magnetic fields effortlessly, filaments perpendicular to the magnetic field gain high density quickly, resulting in a ratio of {\LxbLy} $> 1.5$ for the high-density filaments at $t$ = 0.3 Myr. The shock compression perpendicular to the magnetic field continues slowly. Consequently, the filaments parallel to the magnetic field enter the high-density regime at $t$ = 0.4 Myr, causing the ratio {\LxbLy} to decrease to about 1.  
However, filaments parallel to the magnetic field are mostly dispersed at $t$ = 0.6 after the dissipation of shock compression. Therefore, the ratio {\LxbLy} increases again. The dispersion of these filaments is also responsible for the sudden decrease in the total mass and length of the high-density filaments after $t$ = 0.5 Myr. One possible reason for the dispersion of these filaments is that the magnetic field lines were pinched by the shock compression, resulting in higher field strength and leading to a higher critical line mass \citep[see Equation 28 in][]{Fiege_2000MNRAS}. Consequently, these filaments disperse as soon as the shock compression dissipates. The stability analysis for individual filaments, both aligned parallel and perpendicular to the magnetic field, is beyond the scope of this paper and can be addressed in future work.
\subsection{The formation of HFS}
\label{sec:fil_formation_dis}
As introduced in Section~\ref{sec:intro_JP}, HFSs are assemblies of filaments that transport molecular gas and dust toward the hub \citep{myers_2009}. Thus hub emerges as the densest region, conducive to MSF \citep[e.g.,][]{Dewangan_2023JApA,Dewangan_2024MNRAS,Seshadri_2024MNRAS}. Previously, this data was utilized by \citet{inoue18} to investigate the origin of the most massive sink particle through the gravitational collapse of a massive filament. However, the present work reveals the presence of a massive sink particle and a massive core exactly at the junction of the filaments, indicating that CCC can lead to the formation of HFSs. At $t$ = 0.3 Myr and before, the average velocity of the longitudinal gas flow along the filaments (see the black arrows for $\vec{V}$ in Figure~\ref{Filaments_with_BVG}) is slower ({\simi}0.3 {\kps}) than the turbulent velocity dispersion (as $\Delta v_t$ {\simi}1.0\,{\kps}). The gas flow is also not influenced by gravity, as $\vec{V}$ are randomly oriented with respect to $\vec{g}$ (see Figure~\ref{Delta_PA_dist_app}). Hence, initially, gas flow along the filaments is driven by turbulence. As time progresses, the total mass of the high-density gas increases, allowing gravity to take control of the system's evolution. This shift is reflected in the relative orientations of $\vec{V}$ and $\vec{g}$ at $t$ = 0.4 Myr and beyond. Gravity induces both the mass accumulation of filaments from the surrounding environment and the transport of accumulated gas and dust towards the gravitational center, leading to the convergence of filaments to form HFS \citep[e.g.,][]{Gomez_2018,Wang_2022ApJ}. Therefore, initially, the system's evolution is driven by turbulence, followed by a transition dominated by gravity. Overall, the formation of HFSs from CCC is a three-step process: Step I - Pre-collision phase: Molecular clouds acquire an inhomogeneous density structure driven by turbulence \citep[e.g.,][]{Federrath_2016MNRAS,Shima_2018PASJ,padoan_2020}, Step II - Shock-compression during collision: Filaments are formed due to the oblique shock effect guided by magnetic fields \citep[e.g.,][]{inoue13,inoue18}, and Step III - Convergence of the filaments: Gravity controls the gas kinematics and makes the filaments converge to form a HFS \citep[e.g.,][]{Wang_2022ApJ}. Therefore, the formation of HFS from CCC is a combined effect of turbulence, shock compression, magnetic fields, and gravity. It is important to note that only supersonic CCCs are capable of shock compression; therefore, our model for the formation of HFS from CCC holds exclusively for supersonic collisions.
Earlier, \citet{inoue13} and \citet{inoue18} tested cases where gravity was not incorporated and found that filaments were produced even in the absence of gravity (see Figure 2 in their works). This indicates that the collision process, along with turbulence and the magnetic field, is sufficient for the filament formation.
However, in the later stages, gravity becomes more significant as density increases and attracts gas to form cores within the filaments. A key difference between this study and that of \citet{inoue13} is the initial spherical cloud geometry, which shapes the filaments into a cone and drives inward flows among them. These inward flows along the filaments merge at the vertex of the cone, leading to the substantial accumulation of high-density gas and the formation of massive star(s).

According to \citet{balfour15,Balfour_2017MNRAS}, gravity plays the dominant role in the formation of a network of filaments from CCC, especially when turbulence is subsonic in the colliding molecular clouds. However, our simulation used supersonic turbulence, consistent with the findings of \citet{Larson_1981MNRAS}. In general, the formation of HFSs aligns with the conceptual frameworks proposed in global hierarchical collapse \citep[GHC by][]{vazquez_2009, vazquez_2017, vazquez_2019} or inertial-inflow \citep[][]{padoan_2020} scenarios. Both of these scenarios emphasize the longitudinal inflow of gas along filaments, yet they are driven by distinct mechanisms. GHC is primarily propelled by gravity \citep[][]{vazquez_2019}, whereas inertial inflow is a result of turbulence \citep[][]{padoan_2020}. According to \citet[][]{padoan_2020}, the inertial-inflow model predicts that the net inflow velocity along the filaments is usually significantly smaller than the turbulent velocity and is not primarily controlled by gravity. The GHC model \citep[][]{vazquez_2009, vazquez_2017, vazquez_2019}, on the other hand, predicts anisotropic gravitational contraction at all scales, which is characterized by longitudinal flow along filaments. However, upon recognizing the distinct roles of turbulence and gravity in our model, it becomes evident that neither scenario alone is sufficient to fully explain the formation of HFSs from CCC. Instead, an interplay between turbulence- and gravity-driven theories emerges alongside the effect of collision.

Although the analysis of projected column density maps and field vectors allows for easy comparison with observational works, it is worth noting that some of the correlations between the filament skeletons and the field vectors discussed in this paper may be influenced by the projection effect and mass-weighted averaging process of the vectors.
Interestingly, the impact of mass-weighted averaging in our case is minimal for several reasons. The density in the compressed layer ($n > 10^4$ cm$^{-3}$) greatly exceeds that of both the ambient medium and the dense cloud components, and we have selected only high-density filaments. The x- and y-components of the velocities are zero outside the compressed layer, thus not affecting the direction of the mass-weighted x-y projected velocity components. Although the initial magnetic field strength was 20 $\mu$G, it increased to over 0.3 mG in the compressed layer. Additionally, the gravitational acceleration in the x-y plane is significantly stronger in the compressed layer compared to other regions. Consequently, mass-weighted averaging has a negligible effect on the direction of the x-y projected $\vec{B}$, $\vec{V}$ and $\vec{g}$.
However, to eliminate the projection effect, a more comprehensive analysis would require the use of a simultaneous filament and core identification tool (such as {\it getsf}) for 3D (position-position-position) datasets. This approach would enable the utilization of $\vec{B}$, $\vec{V}$, $\vec{g}$, and their relative orientations for filaments detected in 3D.

\subsection{Difficulties in observational detection of CCC}
\label{sec:CCC_detect_JP}
The primary challenge in observing CCC lies in the angle ($\theta_\mathrm{col}$) of the relative motion of the clouds to the line of sight \citep[see][for details]{fukuia_2018, fujita21}. When $\theta_\mathrm{col}$ is 90$^{\circ}$, colliding clouds become indistinguishable \citep[e.g.,][]{takahira_2014,Priestley2021}, making it impossible to test for CCC. Notably, even under the optimal condition for observing CCC, i.e., at $\theta_\mathrm{col}$ = 0$^{\circ}$, our study reveals that within a timescale, $t_{\mathrm{col}}$ = ${\large{\frac{2R}{v_\mathrm{col}} = \frac{3\,\mathrm{pc}}{10\,\mathrm{km\,s}^{-1}} \approx 0.3}}$ Myr, the smaller cloud will be entirely transformed into the intermediate velocity compressed layer, thereby reducing the velocity separation between the cloud components. 
According to Larson's law, the velocity dispersion of molecular clouds is proportional to $R^{0.5}$ \citep{Larson_1981MNRAS, Ward-Thompson_2015}. Therefore, the FWHM for a cloud with $R = 10$ pc is about 7 {\kps}.
The fact that the velocity dispersion of molecular clouds increases with their size makes it difficult for observers to distinguish between the velocity of the shock-compressed layer and the inherent velocity distribution of the larger cloud component. Therefore, even if a cloud of $R = 1.5$ pc collides with another cloud of $R = 10$ pc at a relative velocity of 10 {\kps}, and it is observed from a perfect viewing angle (i.e., $\theta_\mathrm{col}$ = 0$^{\circ}$), these two cloud components will be indistinguishable after a timescale of about 0.3 Myr irrespective of the velocity resolution of the molecular line data. Furthermore, when colliding clouds exhibit significant size differences, the collision process will result in a minute decrease in the integrated intensity map at the collision site for the larger cloud component. Hence, it is challenging to detect this subtle decrease in emission due to limited instrumental sensitivity. A rough estimate for the minimum detectable cavity size for the SEDIGISM $^{13}$CO(2--1) data can be achieved using the relation, $N(\mathrm{H}_2)$ $\sim10^{21}$ cm$^{-2}$ (K km s$^{-1}$)$^{-1}$ from the studies of \citet{Schuller_2016M, Schuller_2017A&A}. For SEDIGISM $^{13}$CO(2--1) data, sensitivity per channel is about 1\,K and the velocity interval is $\sim$0.25 {\kps} \citep{Schuller2021}. Now, considering a cloud of radius $R = 10$ pc with a velocity distribution spanning about 18 {\kps} (equivalent to an extension of 6$\sigma$ for a FWHM of about 7 {\kps}), a cavity of at least 2 pc is required to detect a 3$\sigma$ (where 1$\sigma$ $\sim$2 K km s$^{-1}$) dip in the moment-0 map for a density of 10$^3$ cm$^{-3}$. Therefore, confirming the complementary distribution and identifying CCC is also challenging for such sites within the timescale of $t_{\mathrm{col}}$. Considering that collisions of molecular clouds with different mass/size scales and similar mass/size scales happen at a similar rate \citep{inoue18}, and $\theta_\mathrm{col}$ can have a wide range of values, it is natural that many CCC sites go undetected. This provides a natural explanation for why the detected CCC sites predominantly consist of clouds of similar sizes \citep{fukui21}.
It is important to note that synthetic molecular line data of CCC will be helpful to search for new observational signatures at $\theta_\mathrm{col}$ = 0$^{\circ}$ when the compressed layer is mixed in velocity space with the larger cloud component.

\subsection{The cone: A mass-collecting machine}
\label{sec:CCC_newsign_JP}
The initial spherical cloud geometry shapes the filaments into a cone, which is best observed at $\theta_\mathrm{col} =$ 90$^{\circ}$. Therefore, the column density maps in the x-z and y-z planes reveal the clearest view of the cone at $t$ = 0.7 Myr in Figure~\ref{combined_fig12}a and \ref{combined_fig12}b, respectively. As discussed in Section~\ref{sec:CCC_detect_JP}, the bridge feature appears in the early phases of the collision, and the detection of complementary distribution can also be challenging if the clouds have a larger size difference. However, the cone shape lasts longer and may be more useful over a much longer time scale (about a few Myr) to test for CCC at $\theta_\mathrm{col} = 90^{\circ}$. The bending of the magnetic field observed in the y-z plane serves as an observational footprint of the CCC or shock compression for an observer at higher values of $\theta_\mathrm{col}$. Interestingly, such curved magnetic field morphology is not observed in the x-z plane (see Figure~\ref{combined_fig12}a) because the initial magnetic field was parallel to the y-axis, i.e., to the line of sight for an observer in that plane. However, an initial magnetic field making some angle with the line of sight for an observer will always show the curved morphology toward the direction of the collision.

As displayed by the projected mass-weighted gas velocity vectors in Figure~\ref{combined_fig12}c, materials on the surface of the cone flow toward its vertex. The total mass of the compressed layer within a cylinder, having its symmetry along the z-axis and radius of 1.5 pc is about 1975 {\Msolar} (including low-density gas and the sink particles). With a density cutoff of 4 $\times$ 10$^{3}$ cm$^{-3}$, the compressed layer contains about 91\% (i.e., {\simi}1805\,{\Msolar}) of its total mass in a total volume of about 1.63 pc$^3$. The high-density layer is divided using cylindrical shells C1, C2, C3, C4, and C5 with radii of 0.3, 0.6, 0.9, 1.2, and 1.5 pc, respectively (see Figure~\ref{combined_fig12}c). Table~\ref{tab4} lists the total mass of the high-density gas, volume fraction, average density, total mass of the sink particles, and the fraction of the total mass for the compressed layer within the cylindrical shells. The vertex of the cone is the densest region of the compressed layer, which is active in star-forming activity as the sink particles are mostly distributed toward it. Hence, the star cluster, including the massive stars, will reside toward the vertex of the cone. Therefore, the distribution of the stars relative to the cone is another observational signature of CCC for higher values of $\theta_\mathrm{col}$. The key feature of this model lies in its capability to accumulate materials rapidly. Initially, the gas of 1805\,{\Msolar} was distributed in a volume of about 36.5 pc$^3$, but it is compressed into a volume of about 1.6 pc$^3$ within a few 0.1 Myr. 
The variations in the dynamic pressure, gravitational acceleration, and the magnetic fields in the compressed layer at $t$ = 0.7 Myr, along the z-direction, are shown in Figures~\ref{combined_fig12}d, \ref{combined_fig12}e, and \ref{combined_fig12}f, respectively. The regions, L1--L6, marked in Figures~\ref{combined_fig12}c,\ref{combined_fig12}d, \ref{combined_fig12}e and \ref{combined_fig12}f, indicate that the strengths of the dynamic pressure, gravitational acceleration, and the magnetic fields increase as we move toward the vertex of the cone from its base.
At the vertex of the cone (i.e., within C1), the total collected mass is about 307\,{\Msolar} within a volume of about 0.13 pc$^3$, making it highly favorable for the formation of massive star(s).

The $PV$ diagrams toward the cone reveal unique features that can hint at the origin of such conical structures from a CCC or shock compression event. The $PV$ diagrams along the y-direction for the regions L1--L6 of the compressed layer are shown in Figures~\ref{combined_fig12_2}a--\ref{combined_fig12_2}f, respectively. The velocity interval between the channels is set to be 0.25 {\kps} (for a y-z-$v_\mathrm{x}$ data cube) and the integration range in the z-direction, $\Delta L_\mathrm{z}$ is 0.25 pc. The positive (i.e., red-shifted) and negative (i.e., blue-shifted) velocity components in the $PV$ diagrams arise from the near- and far-sides of the cone, respectively, due to the projection effect. The gray curves on the $PV$ diagrams represent a constant gas velocity of approximately 4.25 km s$^{-1}$ for a cone with an angle of about 52$^\circ$. The deviation of the $PV$ diagrams from our simulated data and the simple analytical model is because the flow velocity is not constant, and the compressed layer has a finite width. This study reveals that the velocity extent increases as one moves from the base of the cone (i.e., from L6) toward the vertex, due to the gravitational attraction of a large amount of gas collected at the vertex, reaching its maximum at L3. At the vertex, as gas from different directions accumulates, the velocity extent decreases due to the conservation of momentum. Altogether, these signatures in the $PV$ diagrams, the curved morphology of the magnetic field, and the position of the stars in the cone can be utilized to test the CCC scenario at $\theta_\mathrm{col} =$ 90$^{\circ}$.

\citet{Fukui_2017PASJ} conducted a detailed analysis of the {\hone} gas in the southeast region of the LMC and confirmed that the region has two velocity components, labeled as L and D, separated by about 50 {\kps}. The authors found complementary distributions with a spatial displacement between them and suggested that they are colliding with a collision timescale of about 2 Myr. They also discovered {\hone} bridge features between them spanning over a kpc size and interpreted that the collision triggered the formation of the young massive cluster R136 as well as other star-forming regions including N159, located at 500 pc south of 30 Dor. Subsequently, \citet{fukui19ex}, \citet{tokuda19ex,Tokuda_2022ApJ}, and \citet{Wong_2022ApJ} conducted high-resolution CO observations with Atacama Large Millimeter/submillimeter Array (ALMA) toward N159E, N159W-S, and N159W-N, and 30 Dor, respectively. The common properties of these clouds include filamentary cone/fan shapes with north-south elongation ranging from a few pc to a few 10 pc, having a vertex at their southern ends. It is remarkable that most active high-mass star formation in each case takes place near the vertex, in addition to scattered star formation along the filaments. These aspects seem to be in accordance with the MHD simulations of \citet{inoue18}, where the initial setup involves the injection of a spherical cloud onto an extended plane-like gas layer, leading to the formation of filaments into a cone/fan-shaped gas distribution. The present paper extends the model of \citet{inoue18} by revealing detailed physical properties of the gas distribution and kinematics and by elucidating how filaments form and merge at the vertex of the cone to drive massive star-forming activity. We expect to see many more cone/fan-shaped Galactic and extragalactic sources, which will contribute to a better understanding of the model.

\begin{table*}
\centering
  \begin{threeparttable}[b]
   \caption[]{The mass distribution of the compressed layer and the sink particles at $t$ = 0.7 Myr.}
   \centering
   \label{tab4}
   \begin{tabular}{cccccc}
 \hline
Cylindrical  &  $M^{\mathrm{gas}}$ & Volume fraction & $n_{\mathrm{avg}}$ & $M^\mathrm{sink}$ & \large{$\frac{M^{\mathrm{gas}} + M^\mathrm{sink}}{M_{\mathrm{total}}\tnote{1}}$} \\
 shell&  ({\Msolar})&  (\%) &  (cm$^{-3}$) & ({\Msolar}) \\
\hline
C1 &  233 &      8.1 &     29571  & 69 & 0.17\\
C2 &  355 &     15.6 &     23561  & 25 & 0.21\\
C3 &  327 &     20.0 &     16915  &  8 & 0.19\\
C4 &  353 &     25.2 &     14432  & -- & 0.19\\
C5 &  436 &     31.1 &     14469  & -- & 0.24\\
\hline
      \end{tabular}
     \begin{tablenotes}
       \item [1] $M_{\rm total}$ represents the combined mass of the gas and sink particles across all cylindrical shells. 

     \end{tablenotes}
  \end{threeparttable}
\end{table*}

\section{Summary and Conclusions}
\label{sec:sum_JP}
We have analyzed the MHD simulation data  of \citet{inoue18} to understand the connection between two theories of MSF: Accretion through filaments to the hub in case of HFSs and CCC. This study includes a detailed investigation of the shock-compressed layer, the detection of filaments and cores, and an exploration of the gas kinematics involved in CCC. The major outcomes of this work are summarized as follows.
\begin{enumerate}
\item A supersonic CCC creates a shock-compressed layer in the interface of the colliding cloud components. Depending on the initial parameters of the colliding clouds, the morphology of the network of filaments and the core mass spectrum can vary \citep{Balfour_2017MNRAS}. However, with supersonic turbulence or high-density inhomogeneity, the shock compression rapidly leads to the formation of filaments and, subsequently, the creation of HFSs, as detected at $t$ = 0.7 Myr. Therefore, we found that CCC can lead to the formation of HFSs.
\item The collision of the spherical cloud with the sea of dense gas shapes the filaments into a cone and drives inward flows among them. These inward flow of filaments merge at the vertex of the cone, rapidly accumulating high-density gas. Consequently, the vertex of the cone becomes favorable for the formation of a massive star(s). The cone functions as a mass-collecting machine, involving a non-gravitational early process of filament formation, followed by gravitational gas attraction to complete the formation of the HFS.
\item Apart from the angle of the relative motion of the clouds to the line of sight, observational detection of CCC faces two major challenges: I. Identifying the two cloud components in the $PV$ space after the timescale $t_{\mathrm{col}}$ and II. Detecting a minute dip in the emission of the larger cloud component to verify complementary distribution if the cloud components have larger size differences, which is equally probable as mentioned in \citet{inoue18}.

\item CCC events at $\theta_\mathrm{col} = 90^{\circ}$ can be confirmed by the position-velocity diagrams presenting gas flow toward the vertex of the cone, which hosts stars/young stellar objects, and by the magnetic field morphology curved toward the direction of collision.

Altogether, this work indicates a strong connection between CCC and the formation of HFSs. Turbulence, shock compression, magnetic fields, and gravity each contribute to the formation of HFSs through CCC. In addition, this work also highlights the challenges in observing CCC signatures and proposes new possible signatures for $\theta_\mathrm{col} = 90^{\circ}$.

\end{enumerate}

\section*{Acknowledgments}
We thank the anonymous referee for his/her constructive comments, which have improved the scientific content of the manuscript.
A. K. M. acknowledges A. Men’shchikov for scientific conversations on {\it getsf}. The research work at Physical Research Laboratory is funded by the Department of Space, Government of India. A. K. M. acknowledges the use of ParamVikram-1000 HPC facility at Physical Research Laboratory, Ahmedabad, for carrying out the computations of the results [in part] reported in this paper. The numerical computations were carried out on the XC30 system at the Center for Computational Astrophysics (CfCA) of the National Astronomical Observatory of Japan. For figures, we have used {\it matplotlib} \citep{Hunter_2007}.

\begin{figure*}
\center
\includegraphics[width= 0.45\textwidth]{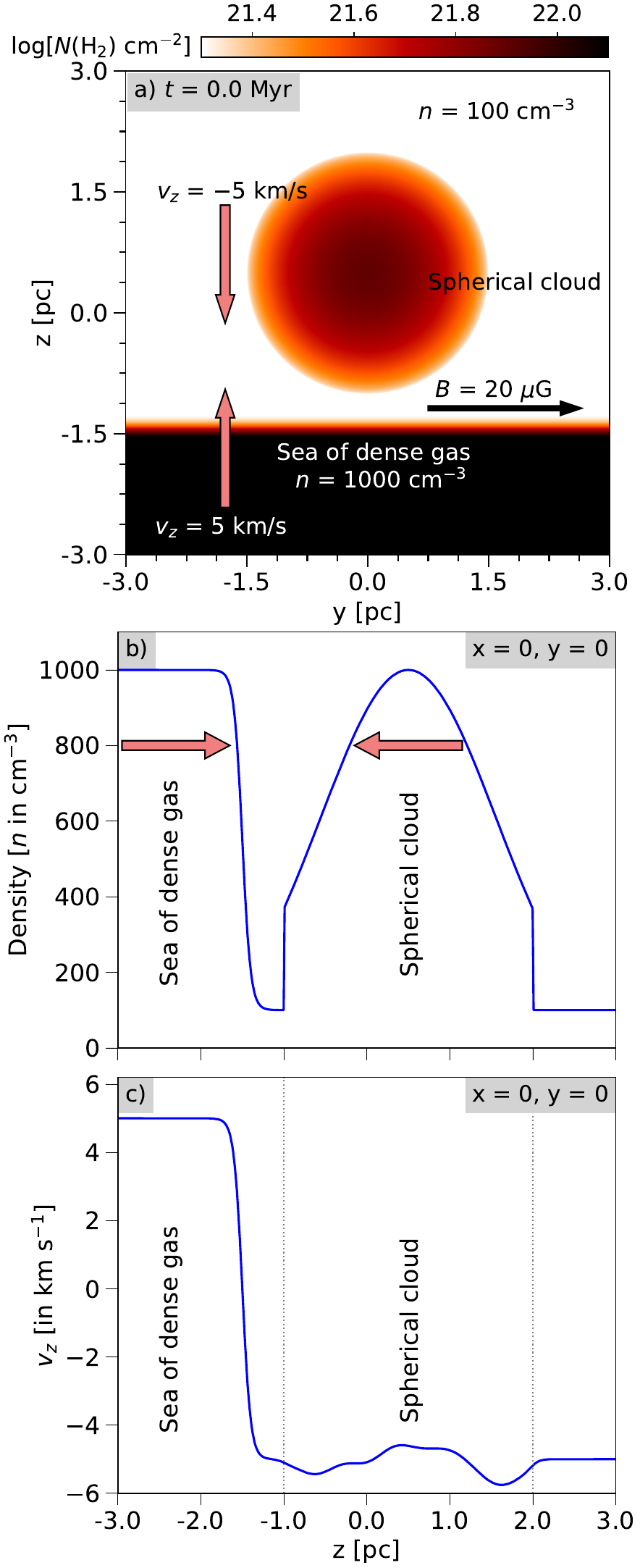}
\caption{An illustration of the initial setup for CCC, similar to \citet{inoue18}: (a) The H$_2$ column density (i.e., $N$(H$_2$)) map in the y-z plane. The arrows in color light-coral point out the directions of movement for the cloud components. The black arrow indicates the direction of the magnetic field. Panels (b) and (c) present the density and z-component of the velocity distribution of the molecular gas along the z-axis, respectively. The spherical cloud component and the sea of dense gas are specified in each panel. In panel ``b,'' the directions of movement of the cloud components are shown with two arrows. In panel ``c,'' the extent of the spherical cloud component is marked with gray dotted lines.}
\label{initial_density_dist}
\end{figure*}

\begin{figure*}
\begin{center}
 \includegraphics[angle=0,width= 0.8\textwidth,trim={0.0cm 0.0cm 0.0cm 0.0cm},clip]{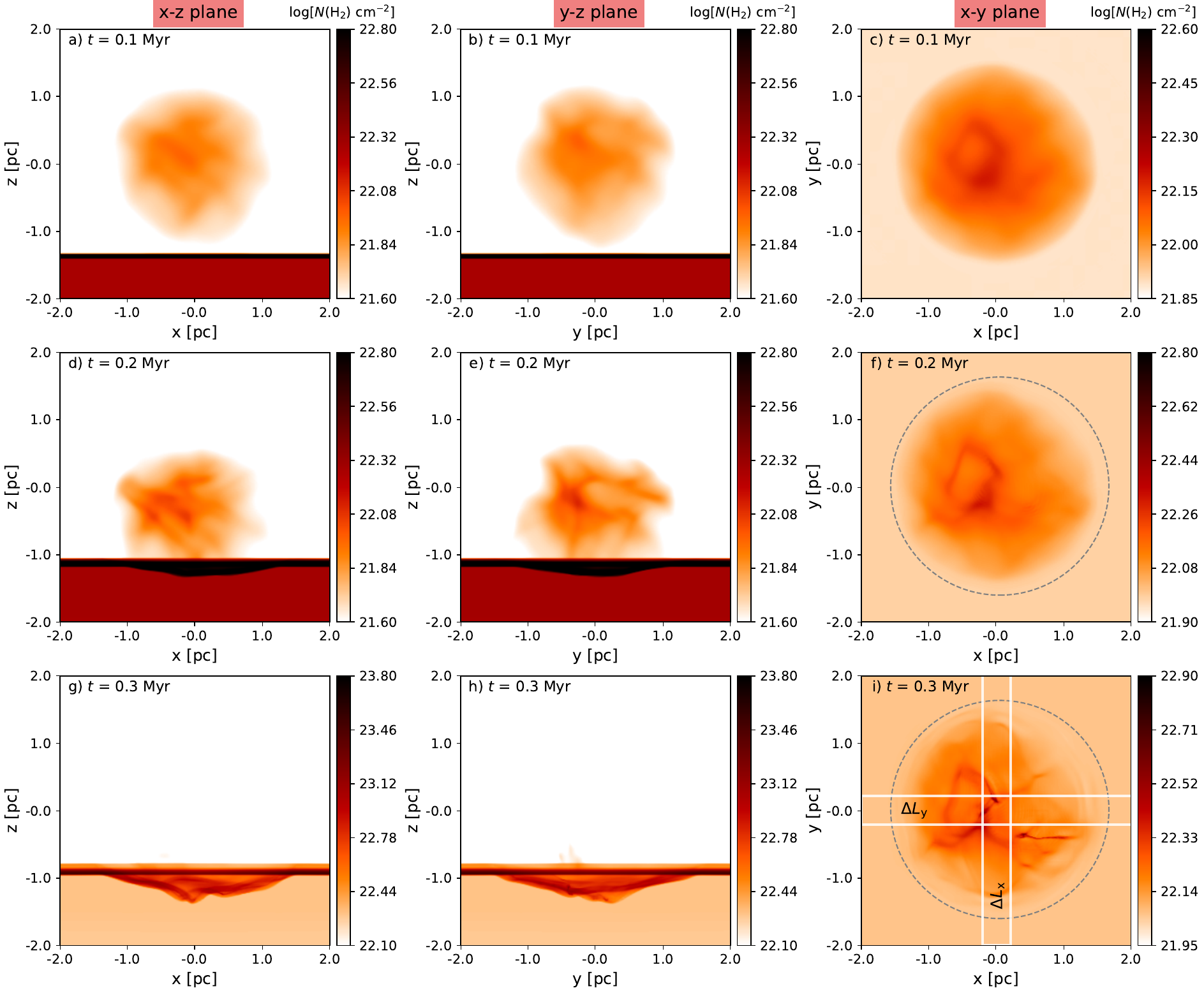}
\caption{Continued.}
\end{center}
\end{figure*}
\addtocounter{figure}{-1}
\begin{figure*}
\begin{center}
 \includegraphics[angle=0,width= 0.8\textwidth,trim={0.0cm 0.0cm 0.0cm 0.0cm},clip]{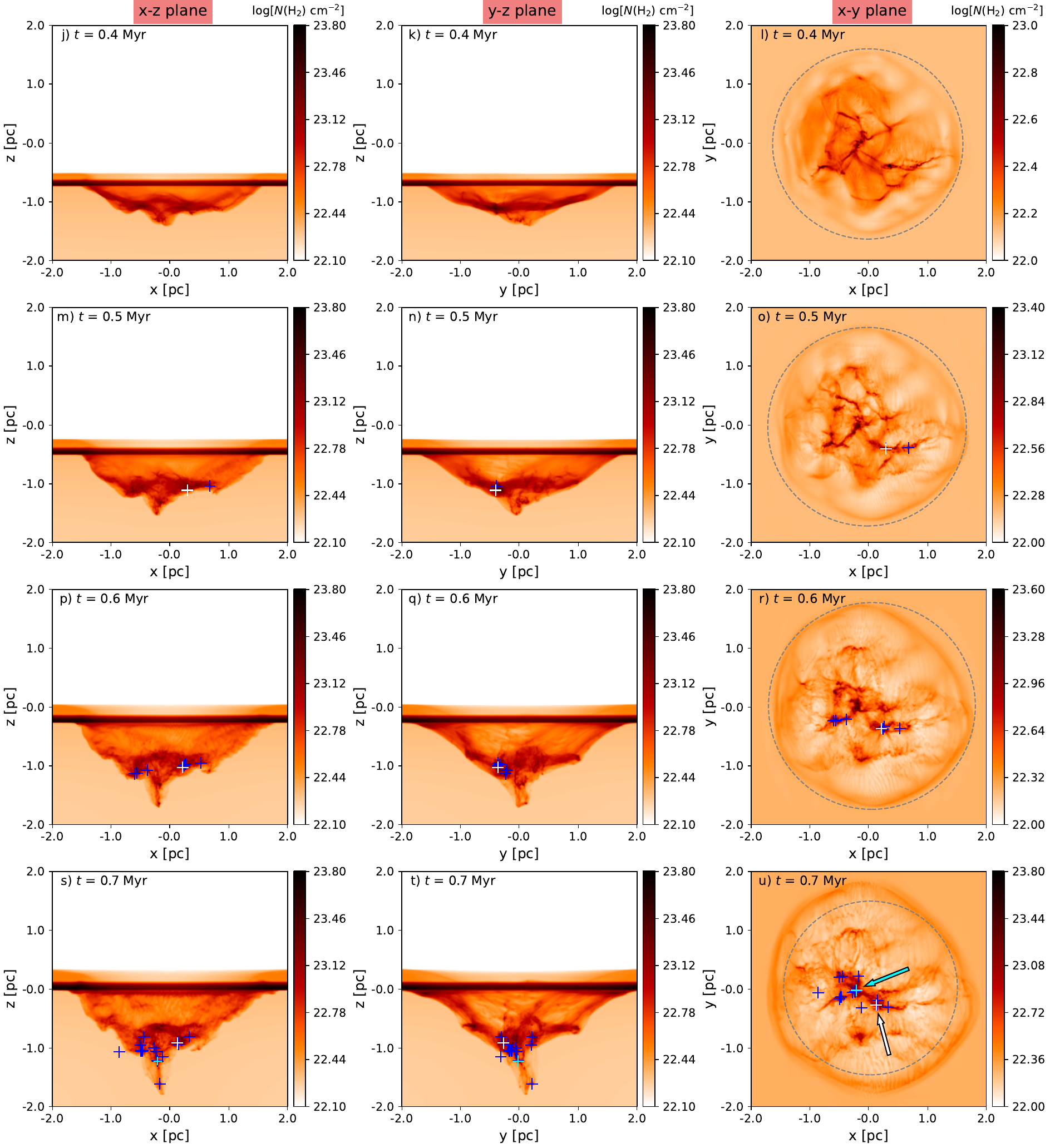}
\caption{The $N$(H$_2$) maps for $t$ = 0.1 to 0.7 Myr in the x-z (at left), y-z (at middle), and x-y (at right) planes, respectively. The plus ($+$) symbols present the positions of the sink particles. The white plus symbol shows the heaviest sink particle. The positions of two massive sink particles at $t$ = 0.7 are marked in white and cyan, with two arrows pointing toward them in panel ``u''. The gray dashed circular regions overlaid on the $N$(H$_2$) maps in the x-y plane from $t$ = 0.2 to 0.7 Myr are zoomed-in in Figure~\ref{NH2_Filaments+cores} with identical color scales. Identical circular regions and color scales from $t$ = 0.3 to 0.7 Myr are presented in Figure~\ref{Filaments_with_BVG}. The pair of horizontal and vertical white lines highlighted on panel ``i'' are utilized to extract the position-velocity ($PV$) diagrams, which are shown in Figure~\ref{PV_withlies}.}
\label{NH2}
\end{center}
\end{figure*}
\begin{figure*}
\begin{center}
 \includegraphics[angle=0,width= 1.0\textwidth,trim={0.0cm 0.0cm 0.0cm 0.0cm},clip]{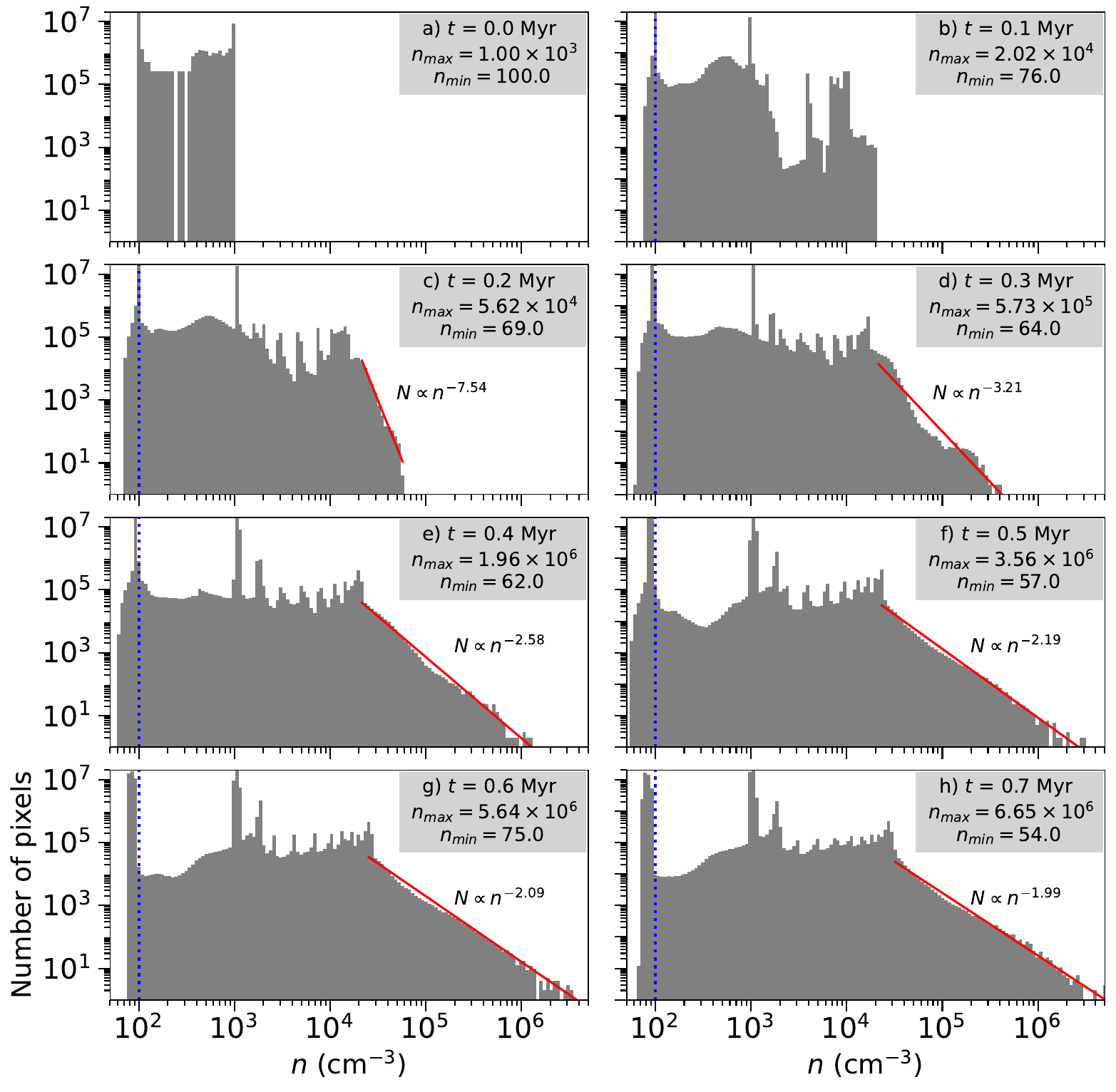}
\caption{Panels (a)--(h) present histograms of H$_2$ density for each pixel within the range of z: $[-2, 2]$ pc, for $t$ = 0.0 to 0.7 Myr, respectively. Each panel also includes the maximum and minimum density values (i.e., $n_{\mathrm{max}}$ and $n_{\mathrm{min}}$, respectively) at the corresponding time. The blue dotted line on the panels ``b''--``h'' is at $n$ = 10$^{2}$ cm$^{-3}$.}
\label{NH2_hist}
\end{center}
\end{figure*}

\begin{figure*}
\begin{center}
 \includegraphics[angle=0,width= 0.8\textwidth,trim={0.0cm 0.0cm 0.0cm 0.0cm},clip]{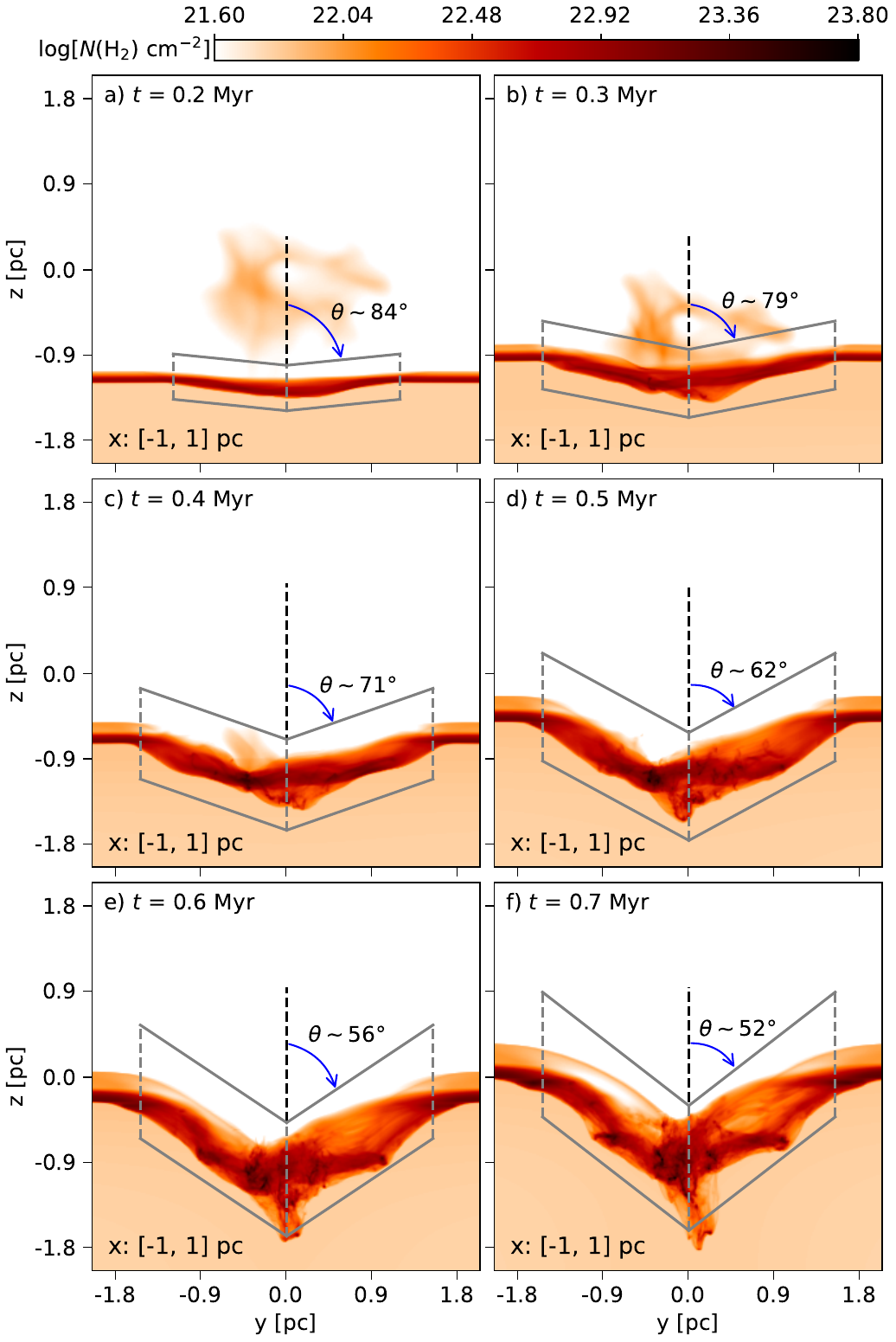}
\caption{Panels (a)--(f) display the $N$(H$_2$) map in the y-z plane for the range of x: $[-1, 1]$ pc. In each panel, the overlaid cone illustrates the conical structure of the shock-compressed layer, and a blue arrow indicates its angle with respect to the z-axis. The angle is also mentioned in each panel.}
\label{cones_fig}
\end{center}
\end{figure*}
\begin{figure}
\begin{center}
 \includegraphics[angle=0,width= 0.5\textwidth,trim={0.0cm 0.0cm 0.0cm 0.0cm},clip]{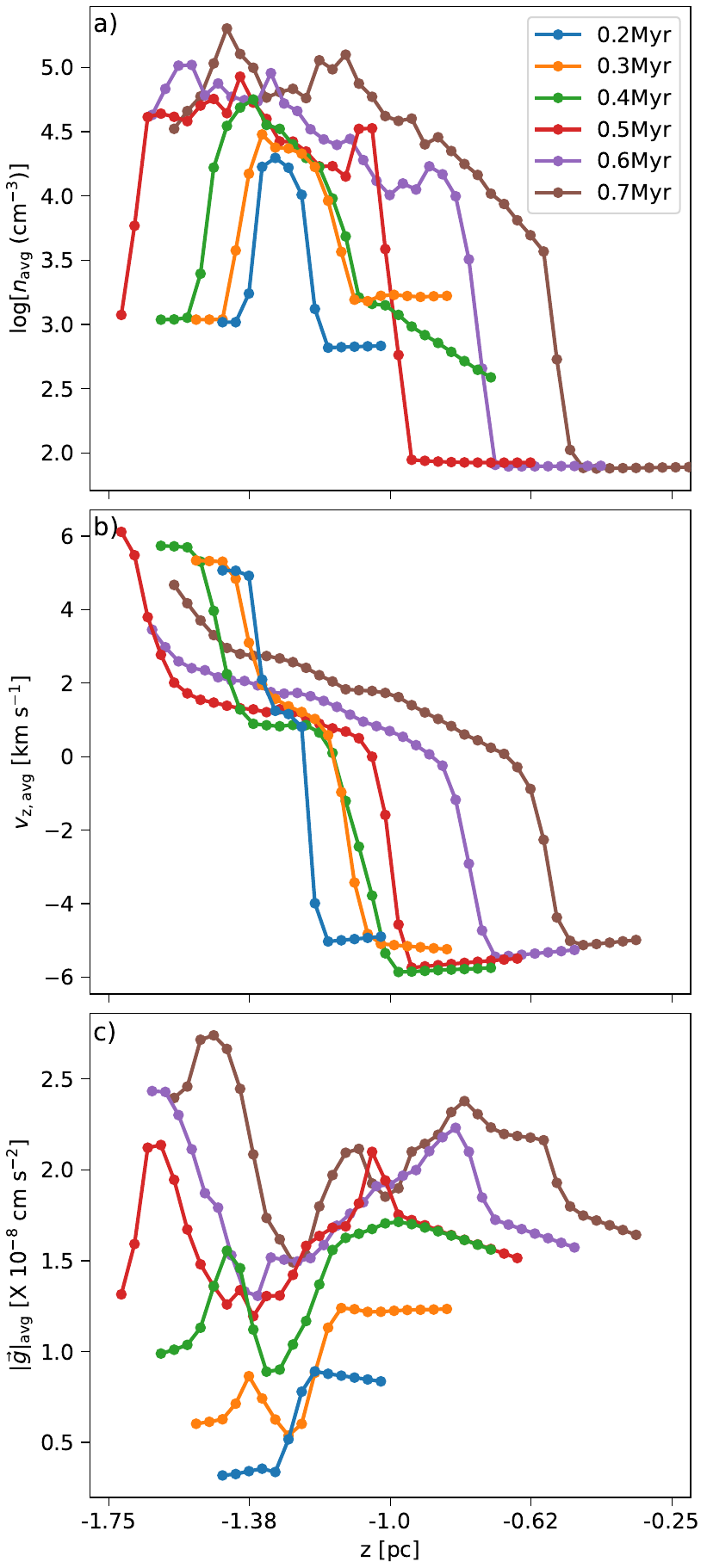}
\caption{The mass-weighted average (a) density and (b) z-component of velocity, and (c) gravitational acceleration of the cone-shaped compressed layers for $t$ = 0.2 to 0.7 Myr as highlighted in Figure~\ref{cones_fig}.}
\label{nvz_dist}
\end{center}
\end{figure}
\begin{figure}
\begin{center}
 \includegraphics[angle=0,width= 0.5\textwidth,trim={0.0cm 0.0cm 0.0cm 0.0cm},clip]{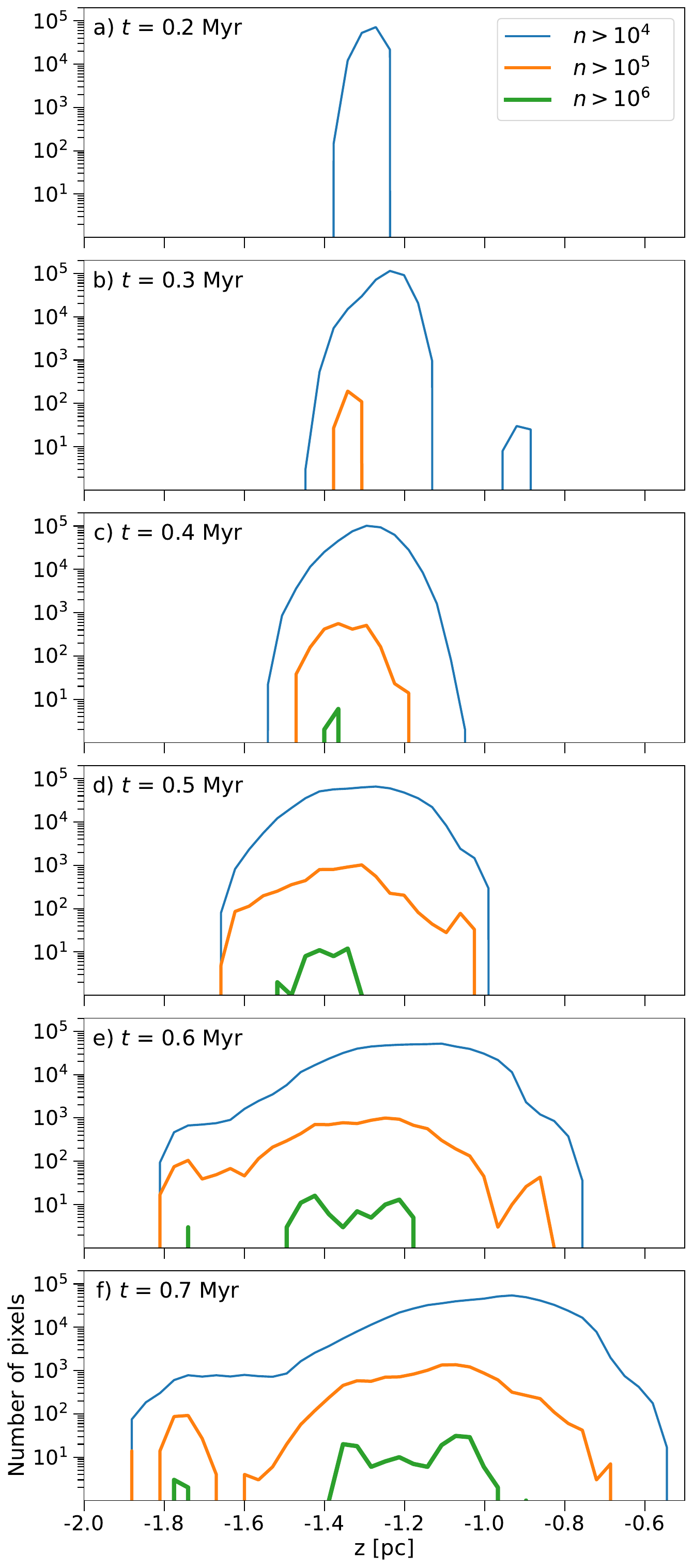}
\caption{Panels (a)--(f) depict the distribution of dense molecular gas along the z-direction for $t$ = 0.2 to 0.7 Myr for the cone-shaped compressed layers shown in Figure~\ref{cones_fig}. However, the z-axis range is extended and made identical for each time. The thinnest, intermediate, and thickest lines represent density thresholds of $n > 10^4$, $n > 10^5$, and $n > 10^6$ cm$^{-3}$, respectively.}
\label{hd_dist}
\end{center}
\end{figure}
\begin{figure}
\begin{center}
 \includegraphics[angle=0,width= 0.5\textwidth,trim={0.0cm 0.0cm 0.0cm 0.0cm},clip]{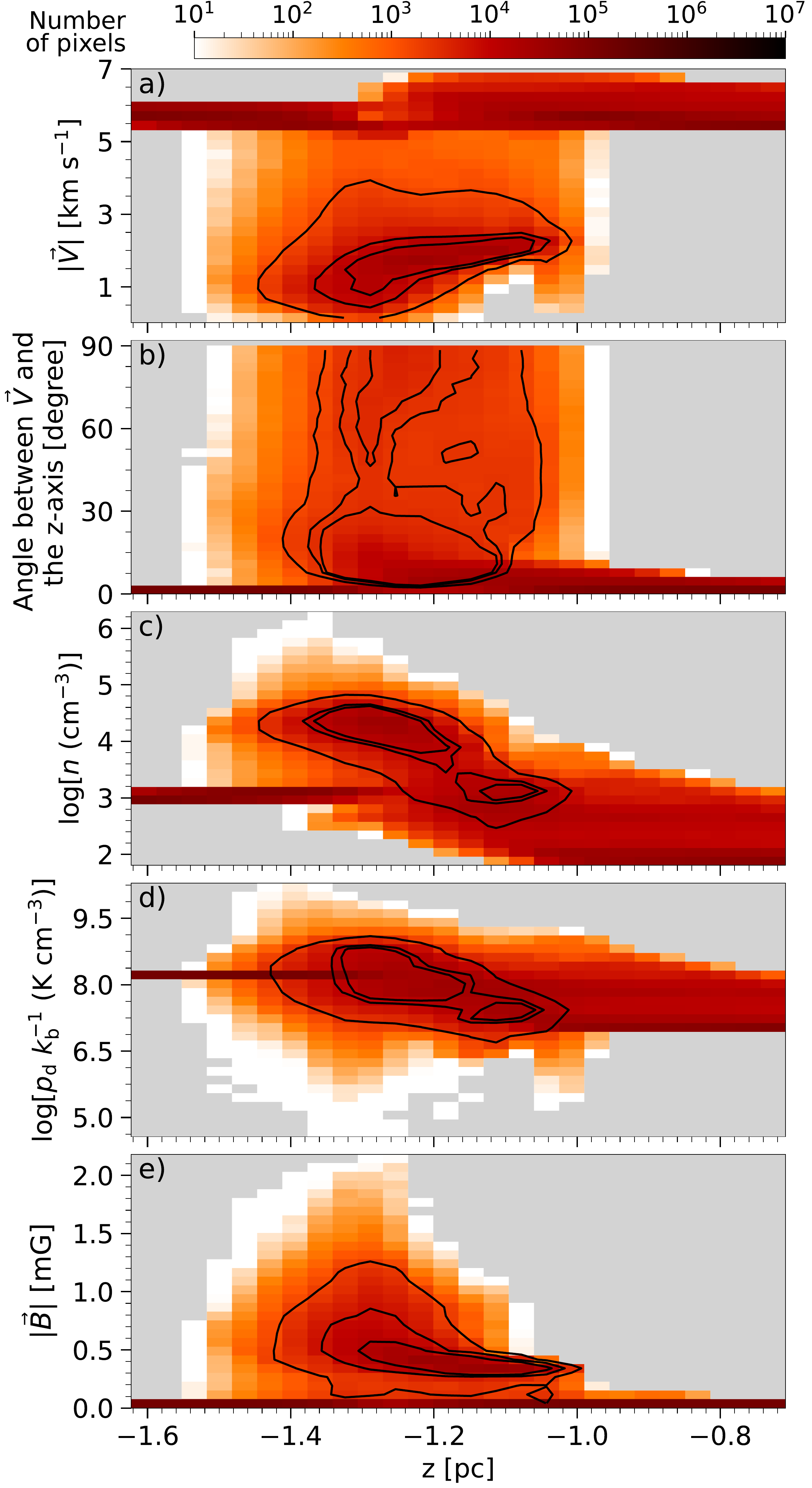}
\caption{The distributions of (a) velocity magnitude ($|\vec{V}|$), (b) angle between $\vec{V}$ and the z-axis, (c) density, (d) dynamic pressure, and (e) magnitude of the magnetic field ($|\vec{B}|$) in the compressed layer at $t$ = 0.4 Myr, corresponding to the conical region shown in Figure~\ref{cones_fig}c. The contours in each panel represent 30\%, 60\%, and 90\% of the total number of pixels, excluding contributions from the initial gas distribution and gravitationally accelerated gas ($|\vec{V}| \geq5$ {\kps} and angle between $\vec{V}$ and the z-axis $\leq5^\circ$).}
\label{04_2dhist}
\end{center}
\end{figure}

%
\begin{figure*}
\begin{center}
 \includegraphics[angle=0,width= 0.9\textwidth,trim={0.0cm 0.0cm 0.0cm 0.0cm},clip]{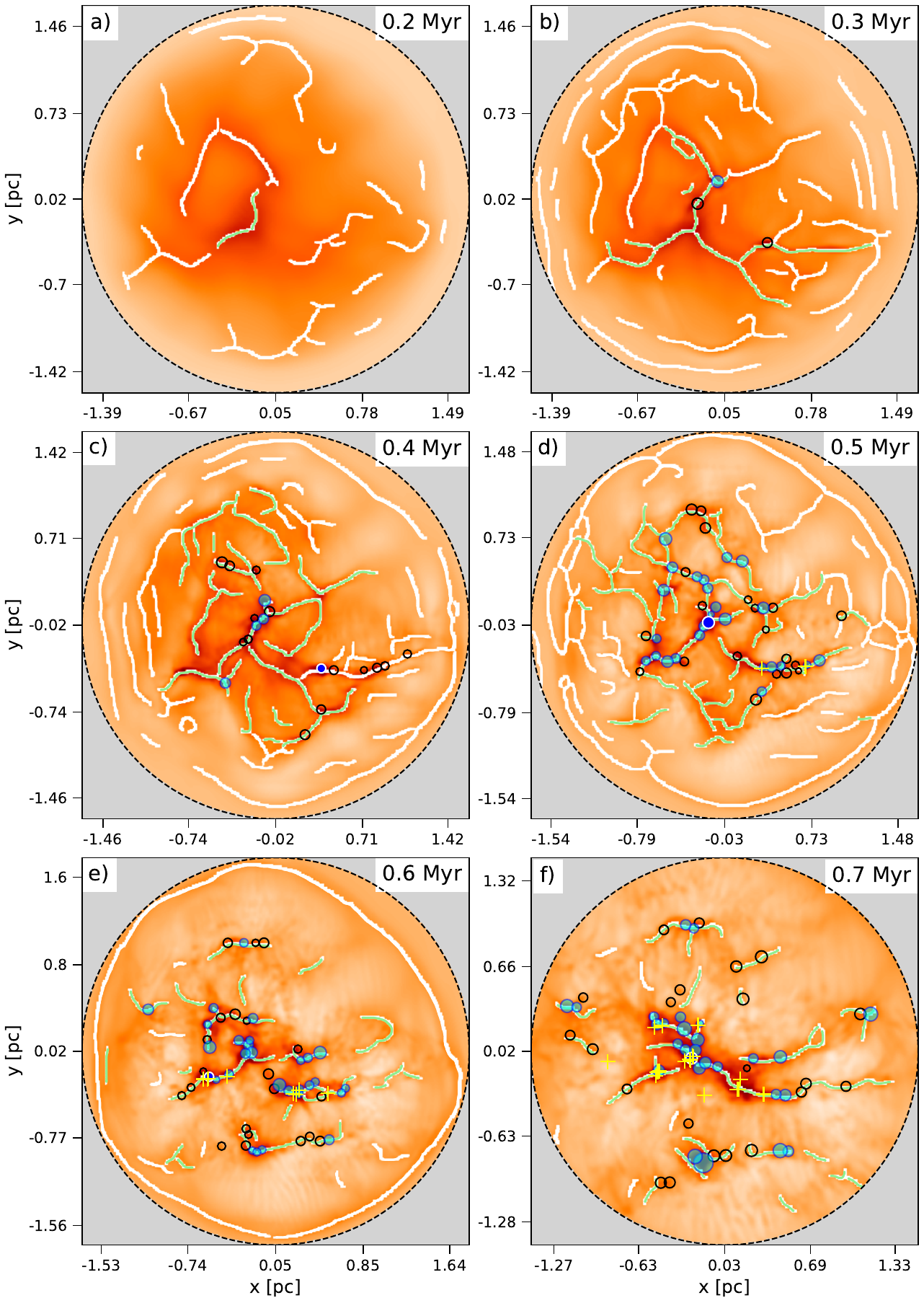}
\caption{Panels (a)--(f) display the distribution of {\it getsf}-identified filament skeletons and cores for $t$ = 0.2 to 0.7 Myr over the $N$(H$_2$) maps, respectively. Filament skeletons are primarily presented in white; however, those with $N(\mathrm{H}_2)^\mathrm{crest}_\mathrm{med} \geq 10^{21.7}$ cm$^{-2}$ are marked in green. Cores with subsolar masses (i.e., $M <$ {\Msolar}) are represented by black circles; those heavier are indicated with filled cyan circles, and the heaviest one is shown with a blue circle outlined in white. The size of the circles represents the actual size of the cores. Position of the sink particles are highlighted with yellow plus symbols.}
\label{NH2_Filaments+cores}
\end{center}
\end{figure*}

\begin{figure*}
\begin{center}
 \includegraphics[angle=0,width= \textwidth,trim={0.0cm 0.0cm 0.0cm 0.0cm},clip]{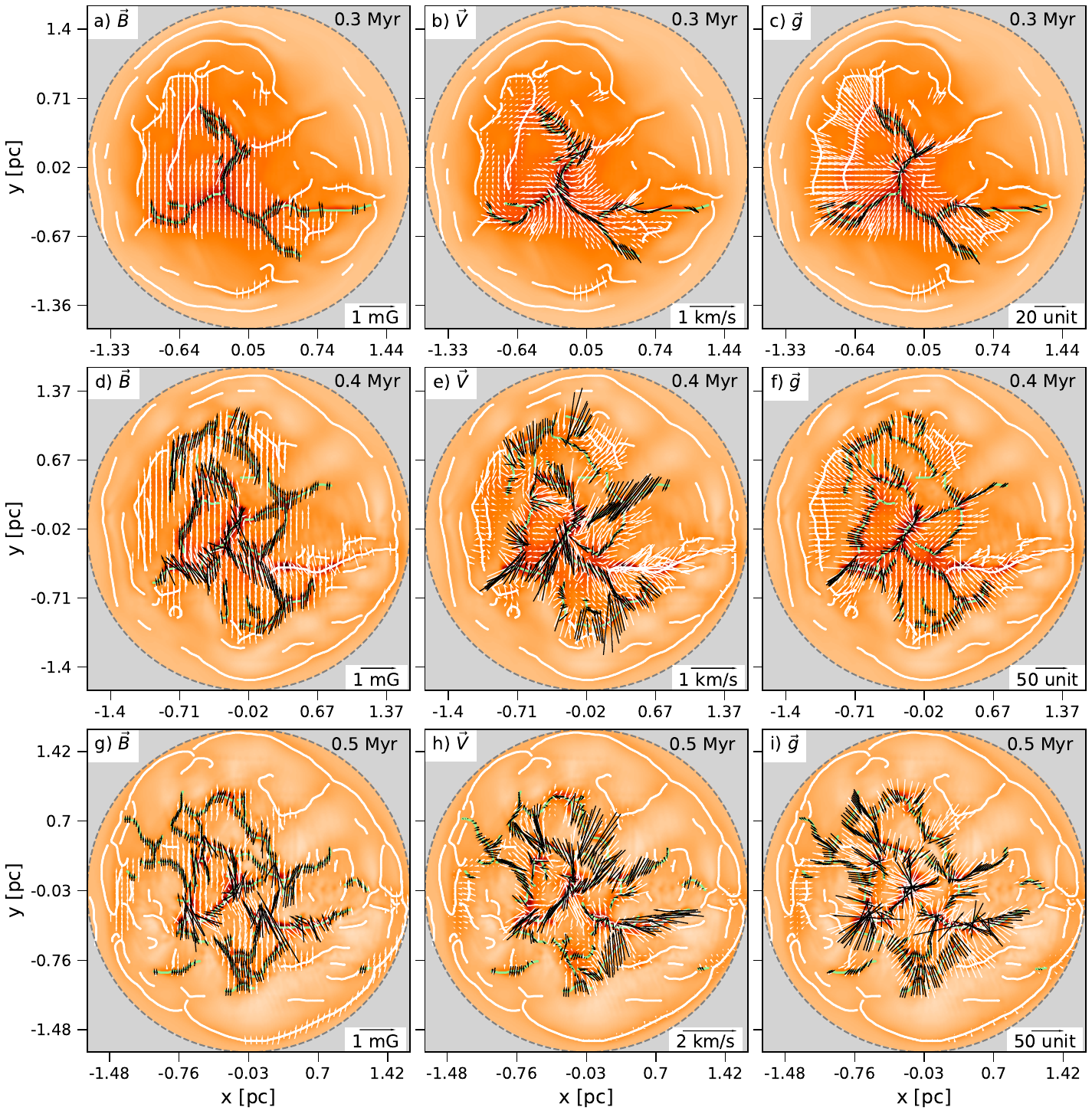}
\caption{Continued.}
\label{Filaments_with_BVG}
\end{center}
\end{figure*}
\addtocounter{figure}{-1}
\begin{figure*}
\begin{center}
 \includegraphics[angle=0,width= \textwidth,trim={0.0cm 0.0cm 0.0cm 0.0cm},clip]{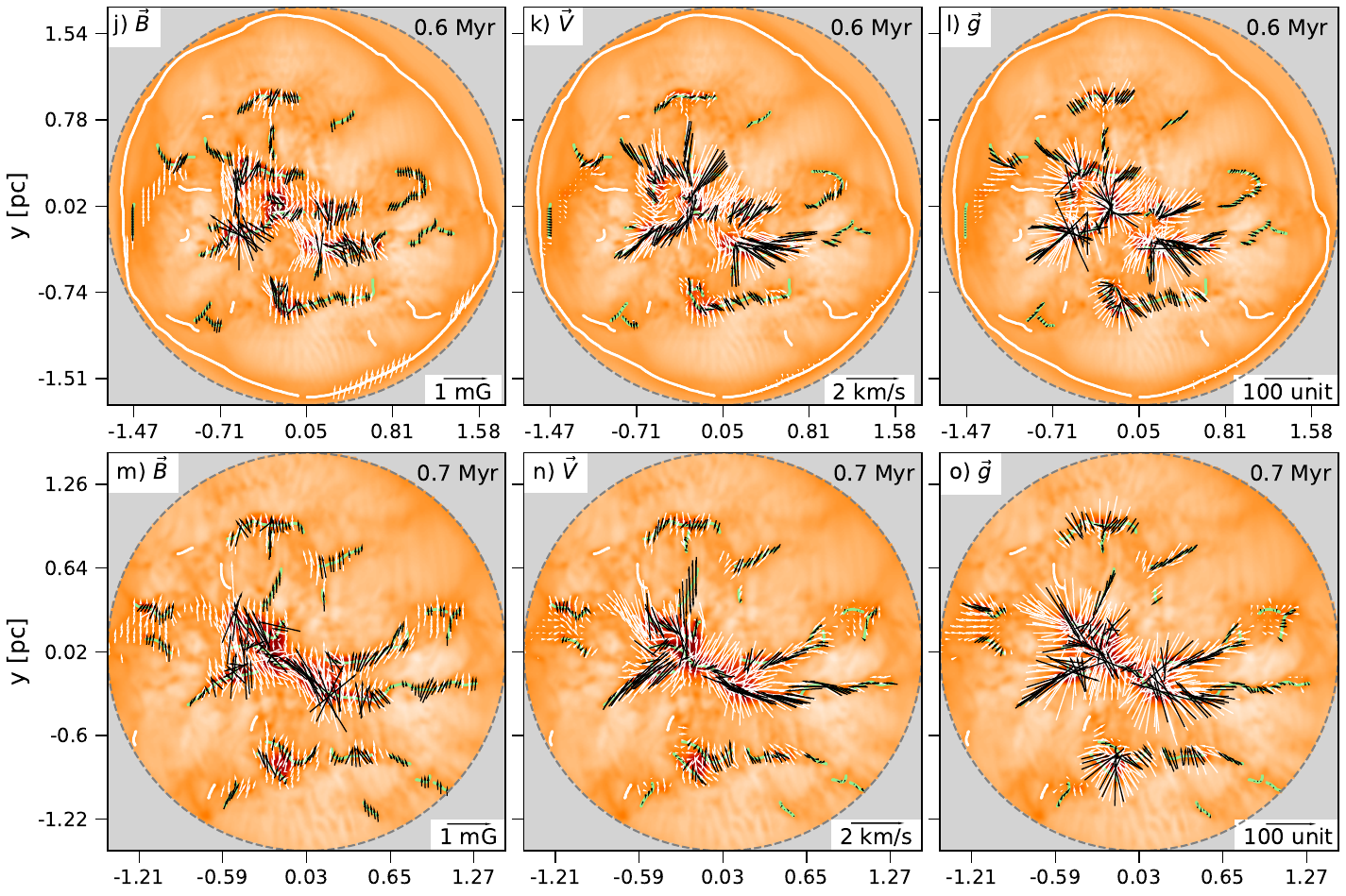}
\caption{The left, middle, and right panels depict the mass-weighted average projected magnetic field, gas velocity, and gravitational field vectors in the x-y plane, respectively. Progressing from top to bottom, the temporal variation of the vectors is illustrated from $t$ = 0.3 to 0.7 Myr. Vectors precisely at high-density filament skeletons (in green) are represented in black, while those in the surrounding regions of the skeletons are depicted in white. Surrounding pixels have column densities above the thresholds of [3, 3, 4, 6, and 6] $\times$ 10$^{21}$ cm$^{-2}$ in the {\it getsf}-identified filament images for $t$ = 0.3, 0.4, 0.5, 0.6, and 0.7 Myr, respectively.}
\label{Filaments_with_BVG}
\end{center}
\end{figure*}

\begin{figure*}
\begin{center}
 \includegraphics[angle=0,width= \textwidth,trim={0.0cm 0.0cm 0.0cm 0.0cm},clip]{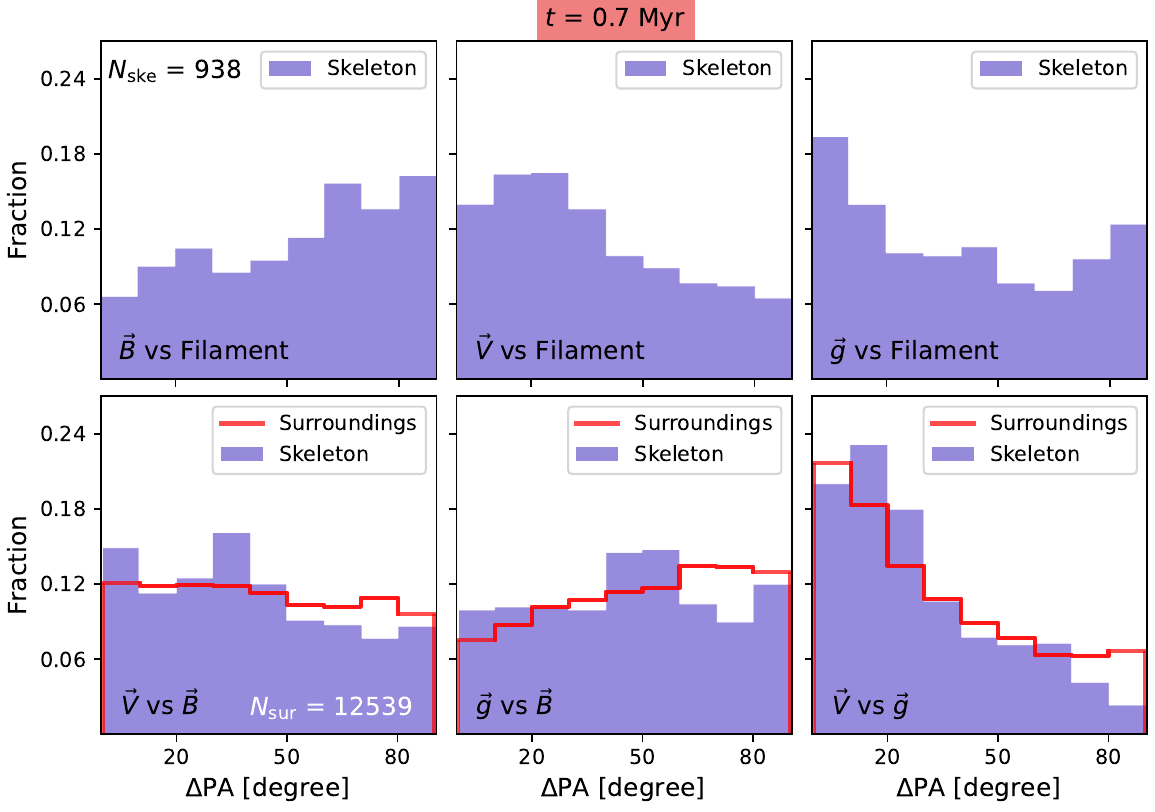}
\caption{Normalized histograms of the relative orientations among the filament skeleton, projected magnetic field, gas velocity, and gravitational field vectors at $t$ = 0.7 Myr. The top row depicts pairs between high-density filament skeletons and the magnetic field, gas velocity, and gravitational field vectors. The bottom row illustrates the relative orientations among the magnetic field, gas velocity, and gravitational field vectors for both the skeletons (in light-slate blue) and their surroundings (in red). The total number of pixels in the filament skeletons ($N_\mathrm{ske}$) and their surroundings ($N_\mathrm{sur}$) are mentioned at each timestep. Figure~\ref{Delta_PA_dist_app} in the Appendix contains similar plots for $t$ = 0.3 to 0.6 Myr.}
\label{Delta_PA_dist}
\end{center}
\end{figure*}
\begin{figure*}
\begin{center}
 \includegraphics[angle=0,width= 0.7\textwidth,trim={0.0cm 0.0cm 0.0cm 0.0cm},clip]{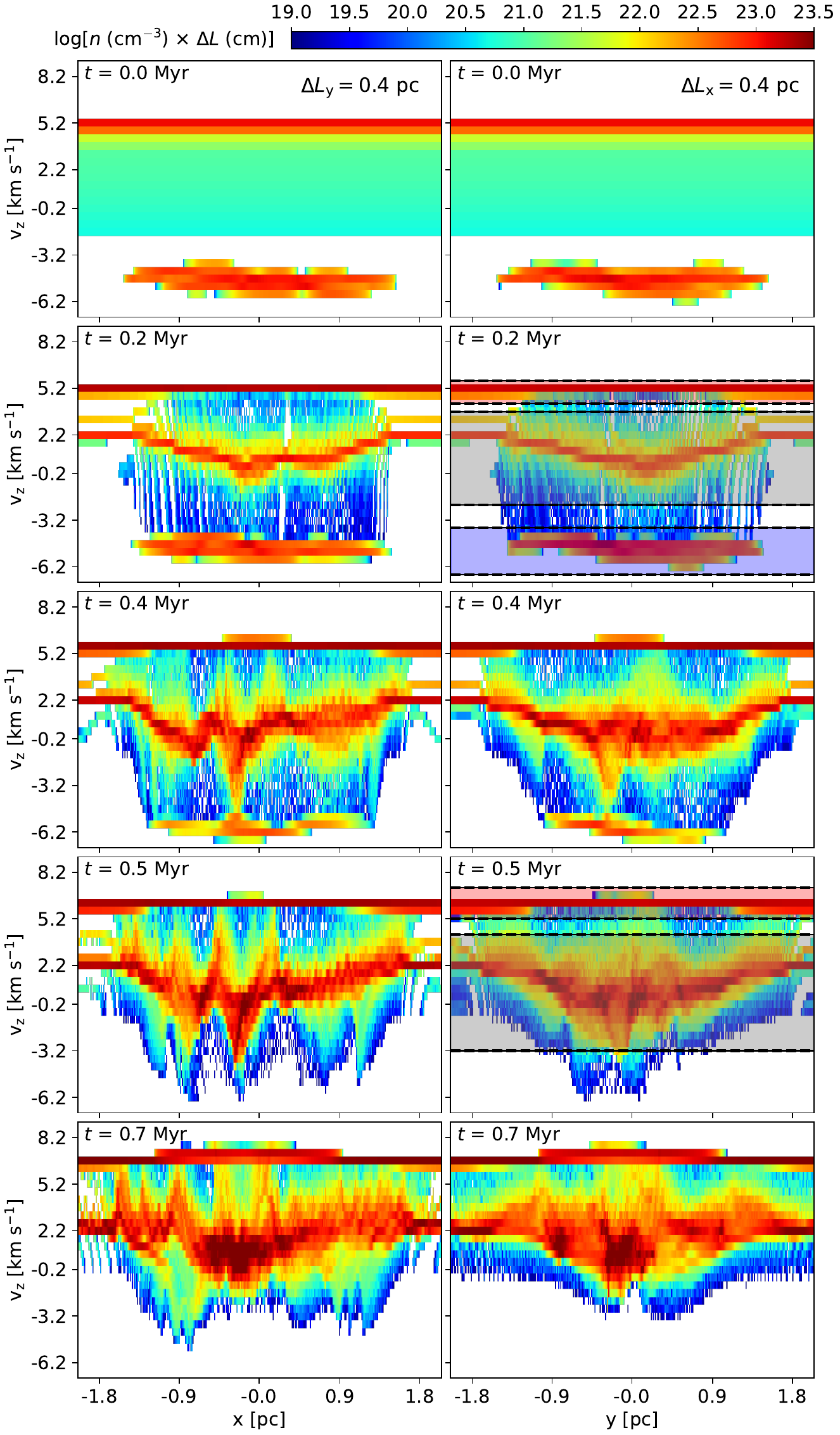}
\caption{The $PV$ diagrams for the regions highlighted in Figure~\ref{NH2}i. The left and right panels are oriented along the x- and y-axes, respectively. Progressing from top to bottom, it covers $t$ = 0, 0.2, 0.4, 0.5, and 0.7 Myr. 
Three velocity ranges are highlighted with shaded regions in blue, gray, and red on the $PV$ diagram along the y-axis at $t$ = 0.2 Myr. Similarly, two velocity ranges are indicated with shaded regions in gray and red on the $PV$ diagram along the y-axis at $t$ = 0.5 Myr. These velocity ranges are utilized to generate integrated density maps, which are presented in Figures~\ref{comp_dist}.}
\label{PV_withlies}
\end{center}
\end{figure*}
\begin{figure*}
\begin{center}
 \includegraphics[angle=0,width= 0.8\textwidth,trim={0.0cm 0.0cm 0.0cm 0.0cm},clip]{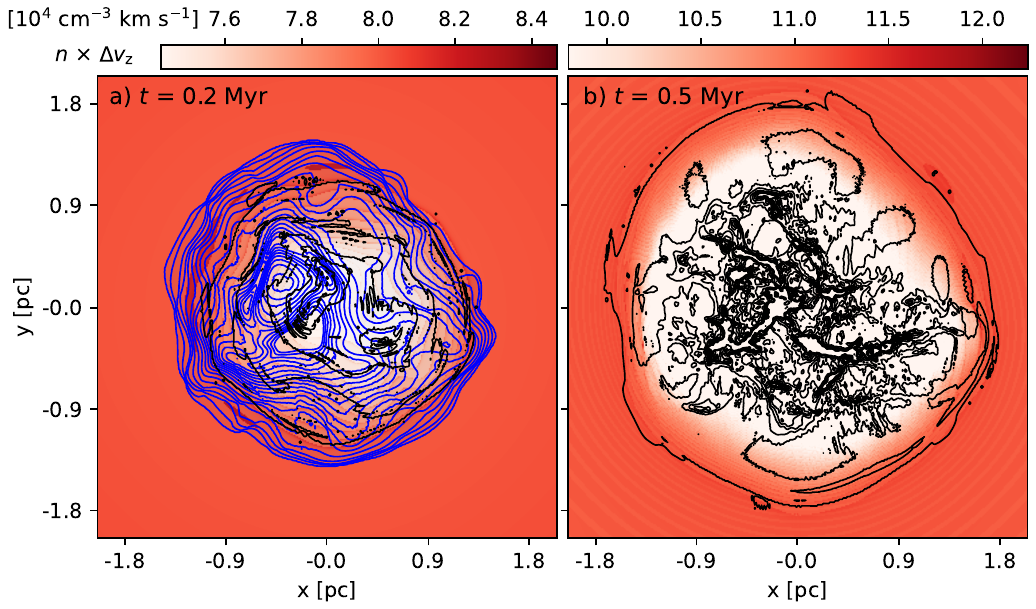}
\caption{The spatial distribution of different velocity components in the x-y plane at $t$ = 0.2 and 0.5 Myr. (a) The red-shifted integrated density image at $t$ = 0.2 Myr is displayed for the velocity range $v_\mathrm{z}$: [4.25, 5.75] {\kps}. The spatial distributions of the blue-shifted component (for $v_\mathrm{z}$: [$-$6.75, $-$3.75] {\kps}) and the compressed layer (for $v_\mathrm{z}$: [$-$2.25, 3.75] {\kps}) are overlaid on the image using blue and black contours, respectively. A total of 20 equispaced contours ranging from 10$^3$ to 10$^5$ cm$^{-3}$ km s$^{-1}$ are shown for the blue-shifted component. Similarly, 10 equispaced contours from 10$^4$ to 10$^5$ cm$^{-3}$ km s$^{-1}$ are shown for the compressed layer. (b) The red-shifted integrated density image at $t$ = 0.5 Myr is presented for the velocity range $v_\mathrm{z}$: [5.25, 7.25] {\kps}. The spatial distribution of the compressed layer (for $v_\mathrm{z}$: [$-$3.25, 4.25] {\kps}) is overlaid on the image using 10 equispaced black contours from 8 $\times$ 10$^2$ to 4 $\times$ 10$^5$ cm$^{-3}$ km s$^{-1}$.
}
\label{comp_dist}
\end{center}
\end{figure*}
\begin{figure*}
\begin{center}
 \includegraphics[angle=0,width= 0.5\textwidth,trim={0.0cm 0.0cm 0.0cm 0.0cm},clip]{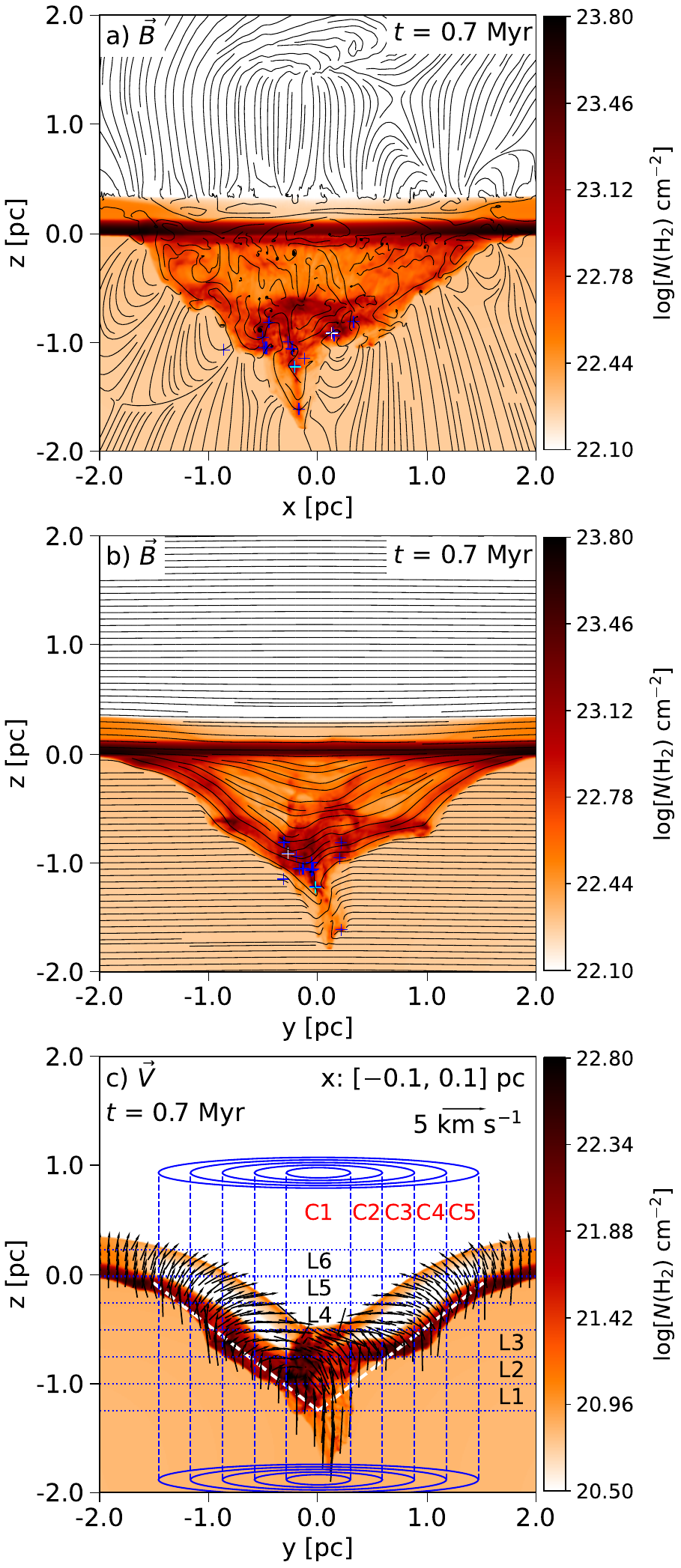}
\includegraphics[angle=0,width= 0.45\textwidth,trim={0.0cm 0.0cm 0.0cm 0.0cm},clip]{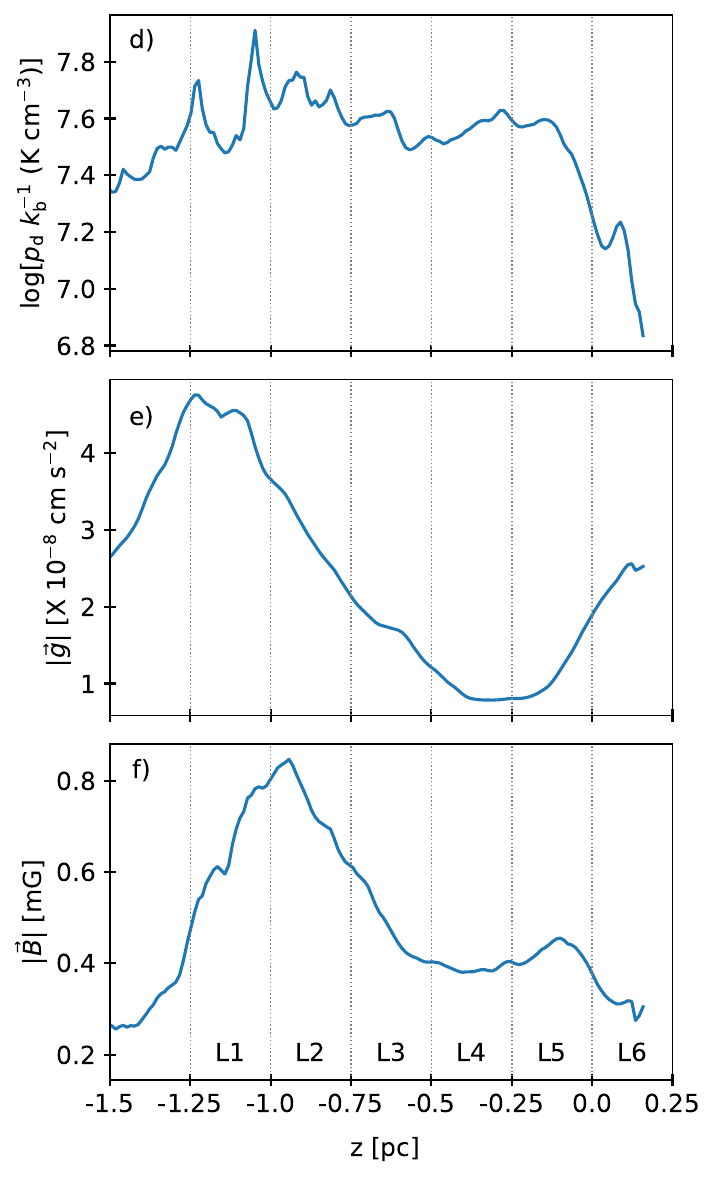}
\caption{The magnetic field streamlines are shown over the (a) x-z and (b) y-z projected column density maps at $t$ = 0.7 Myr. The plus symbols are sink particles, identical to Figure~\ref{NH2}s and \ref{NH2}t, respectively. (c) The y-z projected mass-weighted average gas velocity vectors for the compressed layer are displayed over the column density map for an integration range of x:[$-$0.1, 0.1] pc. The horizontal cuts, L1--L6 (having a width, $\Delta L_\mathrm{z}$ = 0.25 pc), are used to extract the $PV$ diagrams shown in Figure~\ref{combined_fig12_2}. The pair of white dashed lines show a cone (with $\theta$ $\sim$52$^\circ$). Panels (d), (e), and (f) show the changes in dynamic pressure, gravitational acceleration, and the magnetic fields in the compressed layer at $t$ = 0.7 Myr. The regions, L1--L6, are marked in the last three panels with gray dotted lines.}
\label{combined_fig12}
\end{center}
\end{figure*}
\begin{figure*}
\begin{center}
\includegraphics[angle=0,width= 0.8\textwidth,trim={0.0cm 0.0cm 0.0cm 0.0cm},clip]{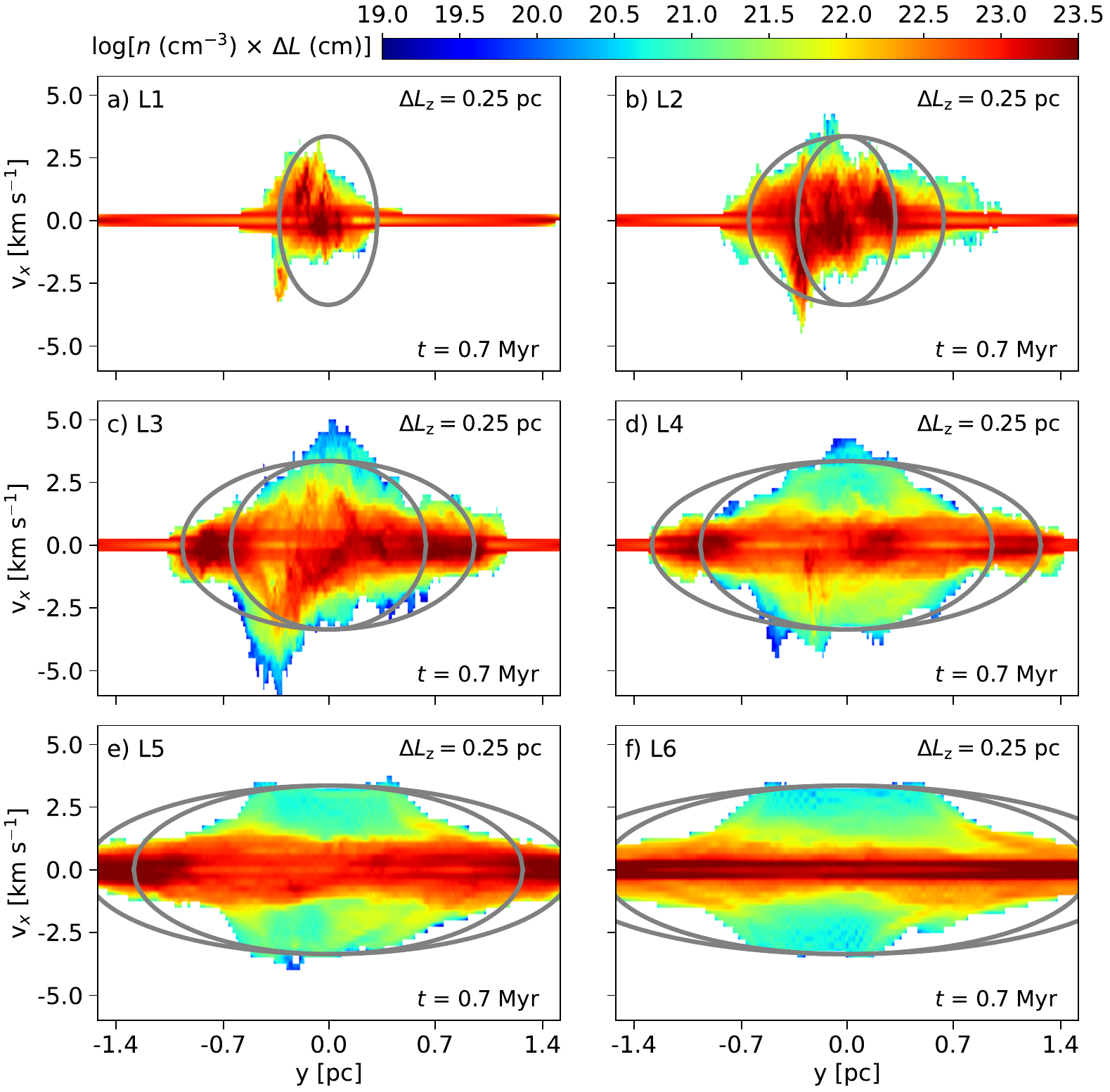}
\caption{Panels (a)--(f) present the $PV$ diagrams extracted for the regions, L1--L6, which are highlighted in Figure~\ref{combined_fig12}c. The gray curves on the $PV$ diagrams present a cone (with $\theta$ $\sim$52$^\circ$) with a constant gas velocity of about 4.25 km s$^{-1}$ directed toward its vertex. The cone is shown by white dashed lines in Figure~\ref{combined_fig12}c.}
\label{combined_fig12_2}
\end{center}
\end{figure*}

\bibliographystyle{aasjournal}
\bibliography{reference}{}

\begin{thebibliography}{}
\expandafter\ifx\csname natexlab\endcsname\relax\def\natexlab#1{#1}\fi
\providecommand{\url}[1]{\href{#1}{#1}}
\providecommand{\dodoi}[1]{doi:~\href{http://doi.org/#1}{\nolinkurl{#1}}}
\providecommand{\doeprint}[1]{\href{http://ascl.net/#1}{\nolinkurl{http://ascl.net/#1}}}
\providecommand{\doarXiv}[1]{\href{https://arxiv.org/abs/#1}{\nolinkurl{https://arxiv.org/abs/#1}}}

\bibitem[{{Abe} {et~al.}(2021){Abe}, {Inoue}, {Inutsuka}, \&
  {Matsumoto}}]{Abe_2021ApJ}
{Abe}, D., {Inoue}, T., {Inutsuka}, S.-i., \& {Matsumoto}, T. 2021, \apj, 916,
  83, \dodoi{10.3847/1538-4357/ac07a1}

\bibitem[{{Anathpindika}(2010)}]{anathpindika_2010}
{Anathpindika}, S.~V. 2010, \mnras, 405, 1431,
  \dodoi{10.1111/j.1365-2966.2010.16541.x}

\bibitem[{{Arreaga-Garc{\'\i}a} {et~al.}(2014){Arreaga-Garc{\'\i}a}, {Klapp},
  \& {Morales}}]{Arreaga-Garcia2014IJAA}
{Arreaga-Garc{\'\i}a}, G., {Klapp}, J., \& {Morales}, J.~S. 2014, International
  Journal of Astronomy and Astrophysics, 4, 192,
  \dodoi{10.4236/ijaa.2014.41018}

\bibitem[{{Arzoumanian} {et~al.}(2011){Arzoumanian}, {Andr{\'e}}, {Didelon},
  {K{\"o}nyves}, {Schneider}, {Men'shchikov}, {Sousbie}, {Zavagno}, {Bontemps},
  {di Francesco}, {Griffin}, {Hennemann}, {Hill}, {Kirk}, {Martin}, {Minier},
  {Molinari}, {Motte}, {Peretto}, {Pezzuto}, {Spinoglio}, {Ward-Thompson},
  {White}, \& {Wilson}}]{Arzoumanian_2011A&A}
{Arzoumanian}, D., {Andr{\'e}}, P., {Didelon}, P., {et~al.} 2011, \aap, 529,
  L6, \dodoi{10.1051/0004-6361/201116596}

\bibitem[{{Auddy} {et~al.}(2018){Auddy}, {Basu}, \& {Kudoh}}]{Auddy_2018MNRAS}
{Auddy}, S., {Basu}, S., \& {Kudoh}, T. 2018, \mnras, 474, 400,
  \dodoi{10.1093/mnras/stx2740}

\bibitem[{{Balfour} {et~al.}(2017){Balfour}, {Whitworth}, \&
  {Hubber}}]{Balfour_2017MNRAS}
{Balfour}, S.~K., {Whitworth}, A.~P., \& {Hubber}, D.~A. 2017, \mnras, 465,
  3483, \dodoi{10.1093/mnras/stw2956}

\bibitem[{{Balfour} {et~al.}(2015){Balfour}, {Whitworth}, {Hubber}, \&
  {Jaffa}}]{balfour15}
{Balfour}, S.~K., {Whitworth}, A.~P., {Hubber}, D.~A., \& {Jaffa}, S.~E. 2015,
  \mnras, 453, 2471, \dodoi{10.1093/mnras/stv1772}

\bibitem[{{Baug} {et~al.}(2016){Baug}, {Dewangan}, {Ojha}, \&
  {Ninan}}]{Baug_2016ApJ}
{Baug}, T., {Dewangan}, L.~K., {Ojha}, D.~K., \& {Ninan}, J.~P. 2016, \apj,
  833, 85, \dodoi{10.3847/1538-4357/833/1/85}

\bibitem[{{Beuther} {et~al.}(2015){Beuther}, {Ragan}, {Johnston}, {Henning},
  {Hacar}, \& {Kainulainen}}]{Beuther_2015A&A}
{Beuther}, H., {Ragan}, S.~E., {Johnston}, K., {et~al.} 2015, \aap, 584, A67,
  \dodoi{10.1051/0004-6361/201527108}

\bibitem[{{Bhadari} {et~al.}(2022){Bhadari}, {Dewangan}, {Ojha}, {Pirogov}, \&
  {Maity}}]{Bhadari_2022ApJ}
{Bhadari}, N.~K., {Dewangan}, L.~K., {Ojha}, D.~K., {Pirogov}, L.~E., \&
  {Maity}, A.~K. 2022, \apj, 930, 169, \dodoi{10.3847/1538-4357/ac65e9}

\bibitem[{{Bhadari} {et~al.}(2020){Bhadari}, {Dewangan}, {Pirogov}, \&
  {Ojha}}]{Bhadari_2020}
{Bhadari}, N.~K., {Dewangan}, L.~K., {Pirogov}, L.~E., \& {Ojha}, D.~K. 2020,
  \apj, 899, 167, \dodoi{10.3847/1538-4357/aba2c6}

\bibitem[{{Cox} {et~al.}(2016){Cox}, {Arzoumanian}, {Andr{\'e}}, {Rygl},
  {Prusti}, {Men'shchikov}, {Royer}, {K{\'o}sp{\'a}l}, {Palmeirim}, {Ribas},
  {K{\"o}nyves}, {Bernard}, {Schneider}, {Bontemps}, {Merin}, {Vavrek}, {Alves
  de Oliveira}, {Didelon}, {Pilbratt}, \& {Waelkens}}]{Cox_2016AA}
{Cox}, N.~L.~J., {Arzoumanian}, D., {Andr{\'e}}, P., {et~al.} 2016, \aap, 590,
  A110, \dodoi{10.1051/0004-6361/201527068}

\bibitem[{{Dewangan}(2017)}]{dewangan_2017}
{Dewangan}, L.~K. 2017, \apj, 837, 44, \dodoi{10.3847/1538-4357/aa5df2}

\bibitem[{{Dewangan} {et~al.}(2024){Dewangan}, {Bhadari}, {Maity}, {Eswaraiah},
  {Sharma}, \& {Jadhav}}]{Dewangan_2024MNRAS}
{Dewangan}, L.~K., {Bhadari}, N.~K., {Maity}, A.~K., {et~al.} 2024, \mnras,
  527, 5895, \dodoi{10.1093/mnras/stad3384}

\bibitem[{{Dewangan} {et~al.}(2023{\natexlab{a}}){Dewangan}, {Bhadari},
  {Maity}, {Pandey}, {Sharma}, {Baug}, \& {Eswaraiah}}]{Dewangan_2023JApA}
---. 2023{\natexlab{a}}, Journal of Astrophysics and Astronomy, 44, 23,
  \dodoi{10.1007/s12036-022-09907-7}

\bibitem[{{Dewangan} {et~al.}(2023{\natexlab{b}}){Dewangan}, {Bhadari},
  {Men'shchikov}, {Chung}, {Devaraj}, {Lee}, {Maity}, \&
  {Baug}}]{Dewangan2023ApJ}
{Dewangan}, L.~K., {Bhadari}, N.~K., {Men'shchikov}, A., {et~al.}
  2023{\natexlab{b}}, \apj, 946, 22, \dodoi{10.3847/1538-4357/acbccc}

\bibitem[{{Dewangan} \& {Ojha}(2017)}]{Dewangan2017}
{Dewangan}, L.~K., \& {Ojha}, D.~K. 2017, \apj, 849, 65,
  \dodoi{10.3847/1538-4357/aa8e00}

\bibitem[{{Dewangan} {et~al.}(2018){Dewangan}, {Ojha}, {Zinchenko}, \&
  {Baug}}]{Dewangan_2018ApJ}
{Dewangan}, L.~K., {Ojha}, D.~K., {Zinchenko}, I., \& {Baug}, T. 2018, \apj,
  861, 19, \dodoi{10.3847/1538-4357/aac6bb}

\bibitem[{{Dobbs} {et~al.}(2015){Dobbs}, {Pringle}, \&
  {Duarte-Cabral}}]{dobbs_2015}
{Dobbs}, C.~L., {Pringle}, J.~E., \& {Duarte-Cabral}, A. 2015, \mnras, 446,
  3608, \dodoi{10.1093/mnras/stu2319}

\bibitem[{{Du}(2021)}]{Du_2021RAA}
{Du}, F. 2021, Research in Astronomy and Astrophysics, 21, 077,
  \dodoi{10.1088/1674-4527/21/3/077}

\bibitem[{{Dunne} {et~al.}(2003){Dunne}, {Eales}, {Ivison}, {Morgan}, \&
  {Edmunds}}]{Dunne2003Nature}
{Dunne}, L., {Eales}, S., {Ivison}, R., {Morgan}, H., \& {Edmunds}, M. 2003,
  \nat, 424, 285, \dodoi{10.1038/nature01792}

\bibitem[{{Enokiya} {et~al.}(2021){Enokiya}, {Ohama}, {Yamada}, {Sano},
  {Fujita}, {Hayashi}, {Tsutsumi}, {Torii}, {Nishimura}, {Konishi}, {Yamamoto},
  {Tachihara}, {Hasegawa}, {Kimura}, {Ogawa}, \& {Fukui}}]{Enokiya2021}
{Enokiya}, R., {Ohama}, A., {Yamada}, R., {et~al.} 2021, \pasj, 73, S256,
  \dodoi{10.1093/pasj/psaa049}

\bibitem[{{Federrath}(2016)}]{Federrath_2016MNRAS}
{Federrath}, C. 2016, \mnras, 457, 375, \dodoi{10.1093/mnras/stv2880}

\bibitem[{{Federrath} {et~al.}(2010){Federrath}, {Roman-Duval}, {Klessen},
  {Schmidt}, \& {Mac Low}}]{Federrath_2010A&A}
{Federrath}, C., {Roman-Duval}, J., {Klessen}, R.~S., {Schmidt}, W., \& {Mac
  Low}, M.~M. 2010, \aap, 512, A81, \dodoi{10.1051/0004-6361/200912437}

\bibitem[{{Fiege} \& {Pudritz}(2000)}]{Fiege_2000MNRAS}
{Fiege}, J.~D., \& {Pudritz}, R.~E. 2000, \mnras, 311, 85,
  \dodoi{10.1046/j.1365-8711.2000.03066.x}

\bibitem[{{Fujita} {et~al.}(2021){Fujita}, {Torii}, {Kuno}, {Nishimura},
  {Umemoto}, {Minamidani}, {Kohno}, {Yamagishi}, {Tosaki}, {Matsuo}, {Tsuda},
  {Enokiya}, {Tachihara}, {Ohama}, {Sano}, {Okawa}, {Hayashi}, {Yoshiike},
  {Tsutsumi}, \& {Fukui}}]{fujita21}
{Fujita}, S., {Torii}, K., {Kuno}, N., {et~al.} 2021, \pasj, 73, S172,
  \dodoi{10.1093/pasj/psz028}

\bibitem[{{Fukui} {et~al.}(2021{\natexlab{a}}){Fukui}, {Habe}, {Inoue},
  {Enokiya}, \& {Tachihara}}]{fukui21}
{Fukui}, Y., {Habe}, A., {Inoue}, T., {Enokiya}, R., \& {Tachihara}, K.
  2021{\natexlab{a}}, \pasj, 73, S1, \dodoi{10.1093/pasj/psaa103}

\bibitem[{{Fukui} {et~al.}(2021{\natexlab{b}}){Fukui}, {Inoue}, {Hayakawa}, \&
  {Torii}}]{Fukui_2021PASJ}
{Fukui}, Y., {Inoue}, T., {Hayakawa}, T., \& {Torii}, K. 2021{\natexlab{b}},
  \pasj, 73, S405, \dodoi{10.1093/pasj/psaa079}

\bibitem[{{Fukui} {et~al.}(2017){Fukui}, {Tsuge}, {Sano}, {Bekki}, {Yozin},
  {Tachihara}, \& {Inoue}}]{Fukui_2017PASJ}
{Fukui}, Y., {Tsuge}, K., {Sano}, H., {et~al.} 2017, \pasj, 69, L5,
  \dodoi{10.1093/pasj/psx032}

\bibitem[{{Fukui} {et~al.}(2014){Fukui}, {Ohama}, {Hanaoka}, {Furukawa},
  {Torii}, {Dawson}, {Mizuno}, {Hasegawa}, {Fukuda}, {Soga}, {Moribe},
  {Kuroda}, {Hayakawa}, {Kawamura}, {Kuwahara}, {Yamamoto}, {Okuda}, {Onishi},
  {Maezawa}, \& {Mizuno}}]{fukui_2014}
{Fukui}, Y., {Ohama}, A., {Hanaoka}, N., {et~al.} 2014, \apj, 780, 36,
  \dodoi{10.1088/0004-637X/780/1/36}

\bibitem[{{Fukui} {et~al.}(2015){Fukui}, {Harada}, {Tokuda}, {Morioka},
  {Onishi}, {Torii}, {Ohama}, {Hattori}, {Nayak}, {Meixner}, {Sewi{\l}o},
  {Indebetouw}, {Kawamura}, {Saigo}, {Yamamoto}, {Tachihara}, {Minamidani},
  {Inoue}, {Madden}, {Galametz}, {Lebouteiller}, {Mizuno}, \&
  {Chen}}]{fukui_2015}
{Fukui}, Y., {Harada}, R., {Tokuda}, K., {et~al.} 2015, \apjl, 807, L4,
  \dodoi{10.1088/2041-8205/807/1/L4}

\bibitem[{{Fukui} {et~al.}(2018{\natexlab{a}}){Fukui}, {Kohno}, {Yokoyama},
  {Nishimura}, {Torii}, {Hattori}, {Sano}, {Ohama}, {Yamamoto}, \&
  {Tachihara}}]{fukui18b}
{Fukui}, Y., {Kohno}, M., {Yokoyama}, K., {et~al.} 2018{\natexlab{a}}, \pasj,
  70, S44, \dodoi{10.1093/pasj/psx144}

\bibitem[{{Fukui} {et~al.}(2018{\natexlab{b}}){Fukui}, {Torii}, {Hattori},
  {Nishimura}, {Ohama}, {Shimajiri}, {Shima}, {Habe}, {Sano}, {Kohno},
  {Yamamoto}, {Tachihara}, \& {Onishi}}]{fukuia_2018}
{Fukui}, Y., {Torii}, K., {Hattori}, Y., {et~al.} 2018{\natexlab{b}}, \apj,
  859, 166, \dodoi{10.3847/1538-4357/aac217}

\bibitem[{{Fukui} {et~al.}(2019){Fukui}, {Tokuda}, {Saigo}, {Harada},
  {Tachihara}, {Tsuge}, {Inoue}, {Torii}, {Nishimura}, {Zahorecz}, {Nayak},
  {Meixner}, {Minamidani}, {Kawamura}, {Mizuno}, {Indebetouw}, {Sewi{\l}o},
  {Madden}, {Galametz}, {Lebouteiller}, {Chen}, \& {Onishi}}]{fukui19ex}
{Fukui}, Y., {Tokuda}, K., {Saigo}, K., {et~al.} 2019, \apj, 886, 14,
  \dodoi{10.3847/1538-4357/ab4900}

\bibitem[{{Furukawa} {et~al.}(2009){Furukawa}, {Dawson}, {Ohama}, {Kawamura},
  {Mizuno}, {Onishi}, \& {Fukui}}]{furukawa_2009}
{Furukawa}, N., {Dawson}, J.~R., {Ohama}, A., {et~al.} 2009, \apjl, 696, L115,
  \dodoi{10.1088/0004-637X/696/2/L115}

\bibitem[{{G{\'o}mez} {et~al.}(2018){G{\'o}mez}, {V{\'a}zquez-Semadeni}, \&
  {Zamora-Avil{\'e}s}}]{Gomez_2018}
{G{\'o}mez}, G.~C., {V{\'a}zquez-Semadeni}, E., \& {Zamora-Avil{\'e}s}, M.
  2018, \mnras, 480, 2939, \dodoi{10.1093/mnras/sty2018}

\bibitem[{{Habe} \& {Ohta}(1992)}]{habe92}
{Habe}, A., \& {Ohta}, K. 1992, \pasj, 44, 203

\bibitem[{{Haworth} {et~al.}(2015){Haworth}, {Shima}, {Tasker}, {Fukui},
  {Torii}, {Dale}, {Takahira}, \& {Habe}}]{haworth_2015}
{Haworth}, T.~J., {Shima}, K., {Tasker}, E.~J., {et~al.} 2015, \mnras, 454,
  1634, \dodoi{10.1093/mnras/stv2068}

\bibitem[{{Hayashi} {et~al.}(2018){Hayashi}, {Sano}, {Enokiya}, {Torii},
  {Hattori}, {Kohno}, {Fujita}, {Nishimura}, {Ohama}, {Yamamoto}, {Tachihara},
  {Hasegawa}, {Kimura}, {Ogawa}, \& {Fukui}}]{hayashi_2018}
{Hayashi}, K., {Sano}, H., {Enokiya}, R., {et~al.} 2018, \pasj, 70, S48,
  \dodoi{10.1093/pasj/psx141}

\bibitem[{{Hennebelle} \& {Falgarone}(2012)}]{Hennebelle_2012A&ARv}
{Hennebelle}, P., \& {Falgarone}, E. 2012, \aapr, 20, 55,
  \dodoi{10.1007/s00159-012-0055-y}

\bibitem[{{Hunter}(2007)}]{Hunter_2007}
{Hunter}, J.~D. 2007, Computing in Science and Engineering, 9, 90,
  \dodoi{10.1109/MCSE.2007.55}

\bibitem[{{Inoue} \& {Fukui}(2013)}]{inoue13}
{Inoue}, T., \& {Fukui}, Y. 2013, \apjl, 774, L31,
  \dodoi{10.1088/2041-8205/774/2/L31}

\bibitem[{{Inoue} {et~al.}(2018){Inoue}, {Hennebelle}, {Fukui}, {Matsumoto},
  {Iwasaki}, \& {Inutsuka}}]{inoue18}
{Inoue}, T., {Hennebelle}, P., {Fukui}, Y., {et~al.} 2018, \pasj, 70, S53,
  \dodoi{10.1093/pasj/psx089}

\bibitem[{{Inutsuka} {et~al.}(2015){Inutsuka}, {Inoue}, {Iwasaki}, \&
  {Hosokawa}}]{Inutsuka_2015A&A}
{Inutsuka}, S.-i., {Inoue}, T., {Iwasaki}, K., \& {Hosokawa}, T. 2015, \aap,
  580, A49, \dodoi{10.1051/0004-6361/201425584}

\bibitem[{{Jog} \& {Ostriker}(1988)}]{Jog_1988ApJ}
{Jog}, C.~J., \& {Ostriker}, J.~P. 1988, \apj, 328, 404, \dodoi{10.1086/166302}

\bibitem[{{Ju{\'a}rez} {et~al.}(2017){Ju{\'a}rez}, {Girart},
  {Zamora-Avil{\'e}s}, {Tang}, {Koch}, {Liu}, {Palau}, {Ballesteros-Paredes},
  {Zhang}, \& {Qiu}}]{Juarez_2017ApJ}
{Ju{\'a}rez}, C., {Girart}, J.~M., {Zamora-Avil{\'e}s}, M., {et~al.} 2017,
  \apj, 844, 44, \dodoi{10.3847/1538-4357/aa78a6}

\bibitem[{{Kim} \& {Ostriker}(2015)}]{Kim_2015ApJ}
{Kim}, C.-G., \& {Ostriker}, E.~C. 2015, \apj, 802, 99,
  \dodoi{10.1088/0004-637X/802/2/99}

\bibitem[{{Klassen} {et~al.}(2017){Klassen}, {Pudritz}, \&
  {Kirk}}]{Klassen_2017MNRAS}
{Klassen}, M., {Pudritz}, R.~E., \& {Kirk}, H. 2017, \mnras, 465, 2254,
  \dodoi{10.1093/mnras/stw2889}

\bibitem[{{Koch} {et~al.}(2022){Koch}, {Tang}, {Ho}, {Hsieh}, {Wang}, {Yen},
  {Duarte-Cabral}, {Peretto}, \& {Su}}]{Koch_2022ApJ}
{Koch}, P.~M., {Tang}, Y.-W., {Ho}, P. T.~P., {et~al.} 2022, \apj, 940, 89,
  \dodoi{10.3847/1538-4357/ac96e3}

\bibitem[{{Krumholz}(2015)}]{Krumholz_2015ASSL}
{Krumholz}, M.~R. 2015, in Astrophysics and Space Science Library, Vol. 412,
  Very Massive Stars in the Local Universe, ed. J.~S. {Vink}, 43,
  \dodoi{10.1007/978-3-319-09596-7_3}

\bibitem[{{Krumholz} {et~al.}(2007){Krumholz}, {Stone}, \&
  {Gardiner}}]{Krumholz_2007}
{Krumholz}, M.~R., {Stone}, J.~M., \& {Gardiner}, T.~A. 2007, \apj, 671, 518,
  \dodoi{10.1086/522665}

\bibitem[{{Kumar} {et~al.}(2020){Kumar}, {Palmeirim}, {Arzoumanian}, \&
  {Inutsuka}}]{Kumar_2020}
{Kumar}, M.~S.~N., {Palmeirim}, P., {Arzoumanian}, D., \& {Inutsuka}, S.~I.
  2020, \aap, 642, A87, \dodoi{10.1051/0004-6361/202038232}

\bibitem[{{Larson}(1981)}]{Larson_1981MNRAS}
{Larson}, R.~B. 1981, \mnras, 194, 809, \dodoi{10.1093/mnras/194.4.809}

\bibitem[{{Liu} {et~al.}(2023){Liu}, {Zhang}, {Koch}, {Liu}, {Li}, {Li},
  {Girart}, {Chen}, {Ching}, {Ho}, {Lai}, {Qiu}, {Rao}, \&
  {Tang}}]{Liu_2023ApJ}
{Liu}, J., {Zhang}, Q., {Koch}, P.~M., {et~al.} 2023, \apj, 945, 160,
  \dodoi{10.3847/1538-4357/acb540}

\bibitem[{{Liu} {et~al.}(2018){Liu}, {Li}, {Juvela}, {Kim}, {Evans}, {Di
  Francesco}, {Liu}, {Yuan}, {Tatematsu}, {Zhang}, {Ward-Thompson}, {Fuller},
  {Goldsmith}, {Koch}, {Sanhueza}, {Ristorcelli}, {Kang}, {Chen}, {Hirano},
  {Wu}, {Sokolov}, {Lee}, {White}, {Wang}, {Eden}, {Li}, {Thompson}, {Pattle},
  {Soam}, {Nasedkin}, {Kim}, {Kim}, {Lai}, {Park}, {Qiu}, {Zhang}, {Alina},
  {Eswaraiah}, {Falgarone}, {Fich}, {Greaves}, {Gu}, {Kwon}, {Li}, {Malinen},
  {Montier}, {Parsons}, {Qin}, {Rawlings}, {Ren}, {Tang}, {Tang}, {Toth},
  {Wang}, {Wouterloot}, {Yi}, \& {Zhang}}]{Liu_2018ApJ}
{Liu}, T., {Li}, P.~S., {Juvela}, M., {et~al.} 2018, \apj, 859, 151,
  \dodoi{10.3847/1538-4357/aac025}

\bibitem[{{Maity} {et~al.}(2023){Maity}, {Dewangan}, {Bhadari}, {Ojha}, {Chen},
  \& {Pandey}}]{Maity_2023MNRAS}
{Maity}, A.~K., {Dewangan}, L.~K., {Bhadari}, N.~K., {et~al.} 2023, \mnras,
  523, 5388, \dodoi{10.1093/mnras/stad1644}

\bibitem[{{Maity} {et~al.}(2022){Maity}, {Dewangan}, {Sano}, {Tachihara},
  {Fukui}, \& {Bhadari}}]{maity_W31}
{Maity}, A.~K., {Dewangan}, L.~K., {Sano}, H., {et~al.} 2022, \apj, 934, 2,
  \dodoi{10.3847/1538-4357/ac7872}

\bibitem[{{Matsumoto}(2007)}]{Matsumoto_2007PASJ}
{Matsumoto}, T. 2007, \pasj, 59, 905, \dodoi{10.1093/pasj/59.5.905}

\bibitem[{{Matsumoto} {et~al.}(2015){Matsumoto}, {Dobashi}, \&
  {Shimoikura}}]{Matsumoto_2015ApJ}
{Matsumoto}, T., {Dobashi}, K., \& {Shimoikura}, T. 2015, \apj, 801, 77,
  \dodoi{10.1088/0004-637X/801/2/77}

\bibitem[{{Men'shchikov}(2021)}]{getsf_2022}
{Men'shchikov}, A. 2021, \aap, 649, A89, \dodoi{10.1051/0004-6361/202039913}

\bibitem[{{Motte} {et~al.}(2018){Motte}, {Bontemps}, \& {Louvet}}]{motte_2018}
{Motte}, F., {Bontemps}, S., \& {Louvet}, F. 2018, \araa, 56, 41,
  \dodoi{10.1146/annurev-astro-091916-055235}

\bibitem[{{Myers}(2009)}]{myers_2009}
{Myers}, P.~C. 2009, \apj, 700, 1609, \dodoi{10.1088/0004-637X/700/2/1609}

\bibitem[{{Navarrete} {et~al.}(2024){Navarrete}, {Pinargote}, \&
  {Banda-Barrag{\'a}n}}]{Navarrete_2024arXiv}
{Navarrete}, S., {Pinargote}, B.~J., \& {Banda-Barrag{\'a}n}, W.~E. 2024, arXiv
  e-prints, arXiv:2404.13250, \dodoi{10.48550/arXiv.2404.13250}

\bibitem[{{Nishimura} {et~al.}(2017){Nishimura}, {Costes}, {Inaba},
  {Tachihara}, {Hattori}, {Kohno}, {Ohama}, {Torii}, {Sano}, {Yamamoto},
  {Hasegawa}, {Kimura}, {Ogawa}, \& {Fukui}}]{nishimura_2017}
{Nishimura}, A., {Costes}, J., {Inaba}, T., {et~al.} 2017, arXiv e-prints,
  arXiv:1706.06002.
\newblock \doarXiv{1706.06002}

\bibitem[{{Nishimura} {et~al.}(2021){Nishimura}, {Fujita}, {Kohno}, {Tsutsumi},
  {Minamidani}, {Torii}, {Umemoto}, {Matsuo}, {Tsuda}, {Kuriki}, {Kuno},
  {Sano}, {Yamamoto}, {Tachihara}, \& {Fukui}}]{Nishimura2021}
{Nishimura}, A., {Fujita}, S., {Kohno}, M., {et~al.} 2021, \pasj, 73, S285,
  \dodoi{10.1093/pasj/psaa083}

\bibitem[{{Orkisz} {et~al.}(2017){Orkisz}, {Pety}, {Gerin}, {Bron},
  {Guzm{\'a}n}, {Bardeau}, {Goicoechea}, {Gratier}, {Le Petit}, {Levrier},
  {Liszt}, {{\"O}berg}, {Peretto}, {Roueff}, {Sievers}, \&
  {Tremblin}}]{Orkisz_2017A&A}
{Orkisz}, J.~H., {Pety}, J., {Gerin}, M., {et~al.} 2017, \aap, 599, A99,
  \dodoi{10.1051/0004-6361/201629220}

\bibitem[{{Padoan} {et~al.}(2016){Padoan}, {Pan}, {Haugb{\o}lle}, \&
  {Nordlund}}]{Padoan_2016ApJ}
{Padoan}, P., {Pan}, L., {Haugb{\o}lle}, T., \& {Nordlund}, {\r{A}}. 2016,
  \apj, 822, 11, \dodoi{10.3847/0004-637X/822/1/11}

\bibitem[{{Padoan} {et~al.}(2020){Padoan}, {Pan}, {Juvela}, {Haugb{\o}lle}, \&
  {Nordlund}}]{padoan_2020}
{Padoan}, P., {Pan}, L., {Juvela}, M., {Haugb{\o}lle}, T., \& {Nordlund},
  {\r{A}}. 2020, \apj, 900, 82, \dodoi{10.3847/1538-4357/abaa47}

\bibitem[{{Park} {et~al.}(2023){Park}, {Currie}, {Thomas}, {Rosolowsky},
  {Dempsey}, {Kim}, {Rigby}, {Su}, {Eden}, {Colombo}, {Parsons}, \&
  {Moore}}]{Park_2023_apj}
{Park}, G., {Currie}, M.~J., {Thomas}, H.~S., {et~al.} 2023, \apjs, 264, 16,
  \dodoi{10.3847/1538-4365/ac9b59}

\bibitem[{{Peretto} {et~al.}(2014){Peretto}, {Fuller}, {Andr{\'e}},
  {Arzoumanian}, {Rivilla}, {Bardeau}, {Duarte Puertas}, {Guzman Fernandez},
  {Lenfestey}, {Li}, {Olguin}, {R{\"o}ck}, {de Villiers}, \&
  {Williams}}]{Peretto_2014}
{Peretto}, N., {Fuller}, G.~A., {Andr{\'e}}, P., {et~al.} 2014, \aap, 561, A83,
  \dodoi{10.1051/0004-6361/201322172}

\bibitem[{{Pillai} {et~al.}(2015){Pillai}, {Kauffmann}, {Tan}, {Goldsmith},
  {Carey}, \& {Menten}}]{Pillai_2015ApJ}
{Pillai}, T., {Kauffmann}, J., {Tan}, J.~C., {et~al.} 2015, \apj, 799, 74,
  \dodoi{10.1088/0004-637X/799/1/74}

\bibitem[{{Pineda} {et~al.}(2023){Pineda}, {Arzoumanian}, {Andre}, {Friesen},
  {Zavagno}, {Clarke}, {Inoue}, {Chen}, {Lee}, {Soler}, \&
  {Kuffmeier}}]{Pineda_2023ASPC}
{Pineda}, J.~E., {Arzoumanian}, D., {Andre}, P., {et~al.} 2023, in Astronomical
  Society of the Pacific Conference Series, Vol. 534, Protostars and Planets
  VII, ed. S.~{Inutsuka}, Y.~{Aikawa}, T.~{Muto}, K.~{Tomida}, \& M.~{Tamura},
  233, \dodoi{10.48550/arXiv.2205.03935}

\bibitem[{{Planck Collaboration} {et~al.}(2016){Planck Collaboration}, {Ade},
  {Aghanim}, {Alves}, {Arnaud}, {Arzoumanian}, {Aumont}, {Baccigalupi},
  {Banday}, {Barreiro}, {Bartolo}, {Battaner}, {Benabed}, {Benoit-L{\'e}vy},
  {Bernard}, {Bern{\'e}}, {Bersanelli}, {Bielewicz}, {Bonaldi}, {Bonavera},
  {Bond}, {Borrill}, {Bouchet}, {Boulanger}, {Bracco}, {Burigana}, {Calabrese},
  {Cardoso}, {Catalano}, {Chamballu}, {Chiang}, {Christensen}, {Clements},
  {Colombi}, {Colombo}, {Combet}, {Couchot}, {Crill}, {Curto}, {Cuttaia},
  {Danese}, {Davies}, {Davis}, {de Bernardis}, {de Rosa}, {de Zotti},
  {Delabrouille}, {Dickinson}, {Diego}, {Donzelli}, {Dor{\'e}}, {Douspis},
  {Ducout}, {Dupac}, {Elsner}, {En{\ss}lin}, {Eriksen}, {Falgarone},
  {Ferri{\`e}re}, {Finelli}, {Forni}, {Frailis}, {Fraisse}, {Franceschi},
  {Frejsel}, {Galeotta}, {Galli}, {Ganga}, {Ghosh}, {Giard},
  {Giraud-H{\'e}raud}, {Gjerl{\o}w}, {Gonz{\'a}lez-Nuevo}, {G{\'o}rski},
  {Gregorio}, {Gruppuso}, {Guillet}, {Hansen}, {Hanson}, {Harrison},
  {Hern{\'a}ndez-Monteagudo}, {Herranz}, {Hildebrandt}, {Hivon}, {Hobson},
  {Holmes}, {Huffenberger}, {Hurier}, {Jaffe}, {Jaffe}, {Jones}, {Juvela},
  {Keskitalo}, {Kisner}, {Knoche}, {Kunz}, {Kurki-Suonio}, {Lagache},
  {Lamarre}, {Lasenby}, {Lawrence}, {Leonardi}, {Levrier}, {Liguori}, {Lilje},
  {Linden-V{\o}rnle}, {L{\'o}pez-Caniego}, {Lubin}, {Mac{\'\i}as-P{\'e}rez},
  {Maffei}, {Mandolesi}, {Mangilli}, {Maris}, {Martin},
  {Mart{\'\i}nez-Gonz{\'a}lez}, {Masi}, {Matarrese}, {Mazzotta}, {Melchiorri},
  {Mendes}, {Mennella}, {Migliaccio}, {Mitra}, {Miville-Desch{\^e}nes},
  {Moneti}, {Montier}, {Morgante}, {Mortlock}, {Munshi}, {Murphy}, {Naselsky},
  {Nati}, {Natoli}, {N{\o}rgaard-Nielsen}, {Noviello}, {Novikov}, {Novikov},
  {Oppermann}, {Pagano}, {Pajot}, {Paladini}, {Paoletti}, {Pasian}, {Perrotta},
  {Pettorino}, {Piacentini}, {Piat}, {Pierpaoli}, {Pietrobon}, {Plaszczynski},
  {Pointecouteau}, {Polenta}, {Pratt}, {Puget}, {Rachen}, {Rebolo}, {Reinecke},
  {Remazeilles}, {Renault}, {Renzi}, {Ricciardi}, {Ristorcelli}, {Rocha},
  {Rosset}, {Rossetti}, {Roudier}, {Rubi{\~n}o-Mart{\'\i}n}, {Rusholme},
  {Sandri}, {Savelainen}, {Savini}, {Scott}, {Soler}, {Stolyarov}, {Sutton},
  {Suur-Uski}, {Sygnet}, {Tauber}, {Terenzi}, {Toffolatti}, {Tomasi},
  {Tristram}, {Tucci}, {Valenziano}, {Valiviita}, {Van Tent}, {Vielva},
  {Villa}, {Wade}, {Wandelt}, {Yvon}, {Zacchei}, \& {Zonca}}]{Planck_2016_feb}
{Planck Collaboration}, {Ade}, P.~A.~R., {Aghanim}, N., {et~al.} 2016, \aap,
  586, A136, \dodoi{10.1051/0004-6361/201425305}

\bibitem[{{Priestley} \& {Whitworth}(2021)}]{Priestley2021}
{Priestley}, F.~D., \& {Whitworth}, A.~P. 2021, \mnras, 506, 775,
  \dodoi{10.1093/mnras/stab1777}

\bibitem[{{Sakre} {et~al.}(2021){Sakre}, {Habe}, {Pettitt}, \&
  {Okamoto}}]{Sakre_2021PASJ}
{Sakre}, N., {Habe}, A., {Pettitt}, A.~R., \& {Okamoto}, T. 2021, \pasj, 73,
  S385, \dodoi{10.1093/pasj/psaa059}

\bibitem[{{Sakre} {et~al.}(2023){Sakre}, {Habe}, {Pettitt}, {Okamoto},
  {Enokiya}, {Fukui}, \& {Hosokawa}}]{Sakre_2023MNRAS}
{Sakre}, N., {Habe}, A., {Pettitt}, A.~R., {et~al.} 2023, \mnras, 522, 4972,
  \dodoi{10.1093/mnras/stad1089}

\bibitem[{{Sano} {et~al.}(2018){Sano}, {Enokiya}, {Hayashi}, {Yamagishi},
  {Saeki}, {Okawa}, {Tsuge}, {Tsutsumi}, {Kohno}, {Hattori}, {Yoshiike},
  {Fujita}, {Nishimura}, {Ohama}, {Tachihara}, {Torii}, {Hasegawa}, {Kimura},
  {Ogawa}, {Wong}, {Braiding}, {Rowell}, {Burton}, \& {Fukui}}]{sano_2018}
{Sano}, H., {Enokiya}, R., {Hayashi}, K., {et~al.} 2018, \pasj, 70, S43,
  \dodoi{10.1093/pasj/psy006}

\bibitem[{{Sano} {et~al.}(2021){Sano}, {Tsuge}, {Tokuda}, {Muraoka},
  {Tachihara}, {Yamane}, {Kohno}, {Fujita}, {Enokiya}, {Rowell}, {Maxted},
  {Filipovi{\'c}}, {Knies}, {Sasaki}, {Onishi}, {Plucinsky}, \&
  {Fukui}}]{Sano_2021PASJ}
{Sano}, H., {Tsuge}, K., {Tokuda}, K., {et~al.} 2021, \pasj, 73, S62,
  \dodoi{10.1093/pasj/psaa045}

\bibitem[{{Schneider} {et~al.}(2012){Schneider}, {Csengeri}, {Hennemann},
  {Motte}, {Didelon}, {Federrath}, {Bontemps}, {Di Francesco}, {Arzoumanian},
  {Minier}, {Andr{\'e}}, {Hill}, {Zavagno}, {Nguyen-Luong}, {Attard},
  {Bernard}, {Elia}, {Fallscheer}, {Griffin}, {Kirk}, {Klessen}, {K{\"o}nyves},
  {Martin}, {Men'shchikov}, {Palmeirim}, {Peretto}, {Pestalozzi}, {Russeil},
  {Sadavoy}, {Sousbie}, {Testi}, {Tremblin}, {Ward-Thompson}, \&
  {White}}]{schenider_2012}
{Schneider}, N., {Csengeri}, T., {Hennemann}, M., {et~al.} 2012, \aap, 540,
  L11, \dodoi{10.1051/0004-6361/201118566}

\bibitem[{{Schneider} {et~al.}(2015){Schneider}, {Csengeri}, {Klessen},
  {Tremblin}, {Ossenkopf}, {Peretto}, {Simon}, {Bontemps}, \&
  {Federrath}}]{Schneider_2015A&A}
{Schneider}, N., {Csengeri}, T., {Klessen}, R.~S., {et~al.} 2015, \aap, 578,
  A29, \dodoi{10.1051/0004-6361/201424375}

\bibitem[{{Schuller} {et~al.}(2016){Schuller}, {Urquhart}, {Bronfman},
  {Csengeri}, {Bontemps}, {Duarte-Cabral}, {Giannetti}, {Ginsburg}, {Henning},
  {Immer}, {Leurini}, {Mattern}, {Menten}, {Molinari}, {Muller},
  {S{\'a}nchez-Monge}, {Schisano}, {Suri}, {Testi}, {Wang}, {Wyrowski}, \&
  {Zavagno}}]{Schuller_2016M}
{Schuller}, F., {Urquhart}, J., {Bronfman}, L., {et~al.} 2016, The Messenger,
  165, 27

\bibitem[{{Schuller} {et~al.}(2017){Schuller}, {Csengeri}, {Urquhart},
  {Duarte-Cabral}, {Barnes}, {Giannetti}, {Hernandez}, {Leurini}, {Mattern},
  {Medina}, {Agurto}, {Azagra}, {Anderson}, {Beltr{\'a}n}, {Beuther},
  {Bontemps}, {Bronfman}, {Dobbs}, {Dumke}, {Finger}, {Ginsburg}, {Gonzalez},
  {Henning}, {Kauffmann}, {Mac-Auliffe}, {Menten}, {Montenegro-Montes},
  {Moore}, {Muller}, {Parra}, {Perez-Beaupuits}, {Pettitt}, {Russeil},
  {S{\'a}nchez-Monge}, {Schilke}, {Schisano}, {Suri}, {Testi}, {Torstensson},
  {Venegas}, {Wang}, {Wienen}, {Wyrowski}, \& {Zavagno}}]{Schuller_2017A&A}
{Schuller}, F., {Csengeri}, T., {Urquhart}, J.~S., {et~al.} 2017, \aap, 601,
  A124, \dodoi{10.1051/0004-6361/201628933}

\bibitem[{{Schuller} {et~al.}(2021){Schuller}, {Urquhart}, {Csengeri},
  {Colombo}, {Duarte-Cabral}, {Mattern}, {Ginsburg}, {Pettitt}, {Wyrowski},
  {Anderson}, {Azagra}, {Barnes}, {Beltran}, {Beuther}, {Billington},
  {Bronfman}, {Cesaroni}, {Dobbs}, {Eden}, {Lee}, {Medina}, {Menten}, {Moore},
  {Montenegro-Montes}, {Ragan}, {Rigby}, {Riener}, {Russeil}, {Schisano},
  {Sanchez-Monge}, {Traficante}, {Zavagno}, {Agurto}, {Bontemps}, {Finger},
  {Giannetti}, {Gonzalez}, {Hernandez}, {Henning}, {Kainulainen}, {Kauffmann},
  {Leurini}, {Lopez}, {Mac-Auliffe}, {Mazumdar}, {Molinari}, {Motte}, {Muller},
  {Nguyen-Luong}, {Parra}, {Perez-Beaupuits}, {Schilke}, {Schneider}, {Suri},
  {Testi}, {Torstensson}, {Veena}, {Venegas}, {Wang}, \&
  {Wienen}}]{Schuller2021}
{Schuller}, F., {Urquhart}, J.~S., {Csengeri}, T., {et~al.} 2021, \mnras, 500,
  3064, \dodoi{10.1093/mnras/staa2369}

\bibitem[{{Seshadri} {et~al.}(2024){Seshadri}, {Vig}, {Ghosh}, \&
  {Ojha}}]{Seshadri_2024MNRAS}
{Seshadri}, A., {Vig}, S., {Ghosh}, S.~K., \& {Ojha}, D.~K. 2024, \mnras, 527,
  4244, \dodoi{10.1093/mnras/stad3385}

\bibitem[{{Shima} {et~al.}(2018){Shima}, {Tasker}, {Federrath}, \&
  {Habe}}]{Shima_2018PASJ}
{Shima}, K., {Tasker}, E.~J., {Federrath}, C., \& {Habe}, A. 2018, \pasj, 70,
  S54, \dodoi{10.1093/pasj/psx124}

\bibitem[{{Stark} \& {Lee}(2005)}]{Stark_2005ApJ}
{Stark}, A.~A., \& {Lee}, Y. 2005, \apjl, 619, L159, \dodoi{10.1086/427936}

\bibitem[{{Takahira} {et~al.}(2014){Takahira}, {Tasker}, \&
  {Habe}}]{takahira_2014}
{Takahira}, K., {Tasker}, E.~J., \& {Habe}, A. 2014, \apj, 792, 63,
  \dodoi{10.1088/0004-637X/792/1/63}

\bibitem[{{Tan} {et~al.}(2014){Tan}, {Beltr{\'a}n}, {Caselli}, {Fontani},
  {Fuente}, {Krumholz}, {McKee}, \& {Stolte}}]{tan14}
{Tan}, J.~C., {Beltr{\'a}n}, M.~T., {Caselli}, P., {et~al.} 2014, in Protostars
  and Planets VI, ed. H.~{Beuther}, R.~S. {Klessen}, C.~P. {Dullemond}, \&
  T.~{Henning}, 149, \dodoi{10.2458/azu\_uapress\_9780816531240-ch007}

\bibitem[{{Tasker} \& {Tan}(2009)}]{tasker_2009}
{Tasker}, E.~J., \& {Tan}, J.~C. 2009, \apj, 700, 358,
  \dodoi{10.1088/0004-637X/700/1/358}

\bibitem[{{Tokuda} {et~al.}(2019){Tokuda}, {Fukui}, {Harada}, {Saigo},
  {Tachihara}, {Tsuge}, {Inoue}, {Torii}, {Nishimura}, {Zahorecz}, {Nayak},
  {Meixner}, {Minamidani}, {Kawamura}, {Mizuno}, {Indebetouw}, {Sewi{\l}o},
  {Madden}, {Galametz}, {Lebouteiller}, {Chen}, \& {Onishi}}]{tokuda19ex}
{Tokuda}, K., {Fukui}, Y., {Harada}, R., {et~al.} 2019, \apj, 886, 15,
  \dodoi{10.3847/1538-4357/ab48ff}

\bibitem[{{Tokuda} {et~al.}(2022){Tokuda}, {Minami}, {Fukui}, {Inoue},
  {Nishioka}, {Tsuge}, {Zahorecz}, {Sano}, {Konishi}, {Rosie Chen},
  {Sewi{\l}o}, {Madden}, {Nayak}, {Saigo}, {Nishimura}, {Tanaka}, {Sawada},
  {Indebetouw}, {Tachihara}, {Kawamura}, \& {Onishi}}]{Tokuda_2022ApJ}
{Tokuda}, K., {Minami}, T., {Fukui}, Y., {et~al.} 2022, \apj, 933, 20,
  \dodoi{10.3847/1538-4357/ac6b3c}

\bibitem[{{Torii} {et~al.}(2011){Torii}, {Enokiya}, {Sano}, {Yoshiike},
  {Hanaoka}, {Ohama}, {Furukawa}, {Dawson}, {Moribe}, {Oishi}, {Nakashima},
  {Okuda}, {Yamamoto}, {Kawamura}, {Mizuno}, {Maezawa}, {Onishi}, {Mizuno}, \&
  {Fukui}}]{torri_2011}
{Torii}, K., {Enokiya}, R., {Sano}, H., {et~al.} 2011, \apj, 738, 46,
  \dodoi{10.1088/0004-637X/738/1/46}

\bibitem[{{Torii} {et~al.}(2015){Torii}, {Hasegawa}, {Hattori}, {Sano},
  {Ohama}, {Yamamoto}, {Tachihara}, {Soga}, {Shimizu}, {Okuda}, {Mizuno},
  {Onishi}, {Mizuno}, \& {Fukui}}]{torri_2015}
{Torii}, K., {Hasegawa}, K., {Hattori}, Y., {et~al.} 2015, \apj, 806, 7,
  \dodoi{10.1088/0004-637X/806/1/7}

\bibitem[{{Torii} {et~al.}(2017){Torii}, {Hattori}, {Hasegawa}, {Ohama},
  {Haworth}, {Shima}, {Habe}, {Tachihara}, {Mizuno}, {Onishi}, {Mizuno}, \&
  {Fukui}}]{torri_2017}
{Torii}, K., {Hattori}, Y., {Hasegawa}, K., {et~al.} 2017, \apj, 835, 142,
  \dodoi{10.3847/1538-4357/835/2/142}

\bibitem[{{Torii} {et~al.}(2021){Torii}, {Hattori}, {Matsuo}, {Fujita},
  {Nishimura}, {Kohno}, {Kuriki}, {Tsuda}, {Minamidani}, {Umemoto}, {Kuno},
  {Yoshiike}, {Ohama}, {Tachihara}, {Fukui}, {Shima}, {Habe}, \&
  {Haworth}}]{torri_2021}
{Torii}, K., {Hattori}, Y., {Matsuo}, M., {et~al.} 2021, \pasj, 73, S368,
  \dodoi{10.1093/pasj/psy098}

\bibitem[{{Trevi{\~n}o-Morales} {et~al.}(2019){Trevi{\~n}o-Morales}, {Fuente},
  {S{\'a}nchez-Monge}, {Kainulainen}, {Didelon}, {Suri}, {Schneider},
  {Ballesteros-Paredes}, {Lee}, {Hennebelle}, {Pilleri},
  {Gonz{\'a}lez-Garc{\'\i}a}, {Kramer}, {Garc{\'\i}a-Burillo}, {Luna},
  {Goicoechea}, {Tremblin}, \& {Geen}}]{trevino19}
{Trevi{\~n}o-Morales}, S.~P., {Fuente}, A., {S{\'a}nchez-Monge}, {\'A}.,
  {et~al.} 2019, \aap, 629, A81, \dodoi{10.1051/0004-6361/201935260}

\bibitem[{{V{\'a}zquez-Semadeni} {et~al.}(2003){V{\'a}zquez-Semadeni},
  {Ballesteros-Paredes}, \& {Klessen}}]{Semadeni_2003ASPC}
{V{\'a}zquez-Semadeni}, E., {Ballesteros-Paredes}, J., \& {Klessen}, R. 2003,
  in Astronomical Society of the Pacific Conference Series, Vol. 287, Galactic
  Star Formation Across the Stellar Mass Spectrum, ed. J.~M. {De Buizer} \&
  N.~S. {van der Bliek}, 81--86, \dodoi{10.48550/arXiv.astro-ph/0206038}

\bibitem[{{V{\'a}zquez-Semadeni} {et~al.}(2009){V{\'a}zquez-Semadeni},
  {G{\'o}mez}, {Jappsen}, {Ballesteros-Paredes}, \& {Klessen}}]{vazquez_2009}
{V{\'a}zquez-Semadeni}, E., {G{\'o}mez}, G.~C., {Jappsen}, A.~K.,
  {Ballesteros-Paredes}, J., \& {Klessen}, R.~S. 2009, \apj, 707, 1023,
  \dodoi{10.1088/0004-637X/707/2/1023}

\bibitem[{{V{\'a}zquez-Semadeni} {et~al.}(2017){V{\'a}zquez-Semadeni},
  {Gonz{\'a}lez-Samaniego}, \& {Col{\'\i}n}}]{vazquez_2017}
{V{\'a}zquez-Semadeni}, E., {Gonz{\'a}lez-Samaniego}, A., \& {Col{\'\i}n}, P.
  2017, \mnras, 467, 1313, \dodoi{10.1093/mnras/stw3229}

\bibitem[{{V{\'a}zquez-Semadeni} {et~al.}(2019){V{\'a}zquez-Semadeni}, {Palau},
  {Ballesteros-Paredes}, {G{\'o}mez}, \& {Zamora-Avil{\'e}s}}]{vazquez_2019}
{V{\'a}zquez-Semadeni}, E., {Palau}, A., {Ballesteros-Paredes}, J.,
  {G{\'o}mez}, G.~C., \& {Zamora-Avil{\'e}s}, M. 2019, \mnras, 490, 3061,
  \dodoi{10.1093/mnras/stz2736}

\bibitem[{{Wang} {et~al.}(2022){Wang}, {Koch}, {Tang}, {Fuller}, {Peretto},
  {Williams}, {Yen}, {Lee}, \& {Chen}}]{Wang_2022ApJ}
{Wang}, J.-W., {Koch}, P.~M., {Tang}, Y.-W., {et~al.} 2022, \apj, 931, 115,
  \dodoi{10.3847/1538-4357/ac6872}

\bibitem[{{Ward-Thompson} \& {Whitworth}(2015)}]{Ward-Thompson_2015}
{Ward-Thompson}, D., \& {Whitworth}, A.~P. 2015, {An Introduction to Star
  Formation}

\bibitem[{{Wong} {et~al.}(2022){Wong}, {Oudshoorn}, {Sofovich}, {Green},
  {Shah}, {Indebetouw}, {Meixner}, {Hacar}, {Nayak}, {Tokuda}, {Bolatto},
  {Chevance}, {De Marchi}, {Fukui}, {Hirschauer}, {Jameson}, {Kalari},
  {Lebouteiller}, {Looney}, {Madden}, {Onishi}, {Roman-Duval}, {Rubio}, \&
  {Tielens}}]{Wong_2022ApJ}
{Wong}, T., {Oudshoorn}, L., {Sofovich}, E., {et~al.} 2022, \apj, 932, 47,
  \dodoi{10.3847/1538-4357/ac723a}

\bibitem[{{Zhou} {et~al.}(2022){Zhou}, {Liu}, {Evans}, {Garay}, {Goldsmith},
  {G{\'o}mez}, {V{\'a}zquez-Semadeni}, {Liu}, {Stutz}, {Wang}, {Juvela}, {He},
  {Li}, {Bronfman}, {Liu}, {Xu}, {Tej}, {Dewangan}, {Li}, {Zhang}, {Zhang},
  {Ren}, {Tatematsu}, {Shing Li}, {Won Lee}, {Baug}, {Qin}, {Wu}, {Peng},
  {Zhang}, {Liu}, {Luo}, {Ge}, {Saha}, {Chakali}, {Zhang}, {Kim},
  {Ristorcelli}, {Shen}, \& {Li}}]{Zhou_2022MNRAS}
{Zhou}, J.-W., {Liu}, T., {Evans}, N.~J., {et~al.} 2022, \mnras, 514, 6038,
  \dodoi{10.1093/mnras/stac1735}

\bibitem[{{Zhou} {et~al.}(2023){Zhou}, {Li}, {Liu}, {Peng}, {Zhang}, {Xu},
  {Zhang}, {Liu}, \& {Li}}]{Zhou_2023MNRAS}
{Zhou}, J.-W., {Li}, S., {Liu}, H.-L., {et~al.} 2023, \mnras, 519, 2391,
  \dodoi{10.1093/mnras/stac3559}

\bibitem[{{Zinnecker} \& {Yorke}(2007)}]{zinnecker07}
{Zinnecker}, H., \& {Yorke}, H.~W. 2007, \araa, 45, 481,
  \dodoi{10.1146/annurev.astro.44.051905.092549}

\end{thebibliography}
\appendix
\section{The $\Delta$PA distributions for $t$ = 0.3 to 0.6 Myr}
\label{sec:DeltaPA_appendix}
\begin{figure*}[b]
\begin{center}
 \includegraphics[angle=0,width= 0.7\textwidth,trim={0.0cm 0.0cm 0.0cm 0.0cm},clip]{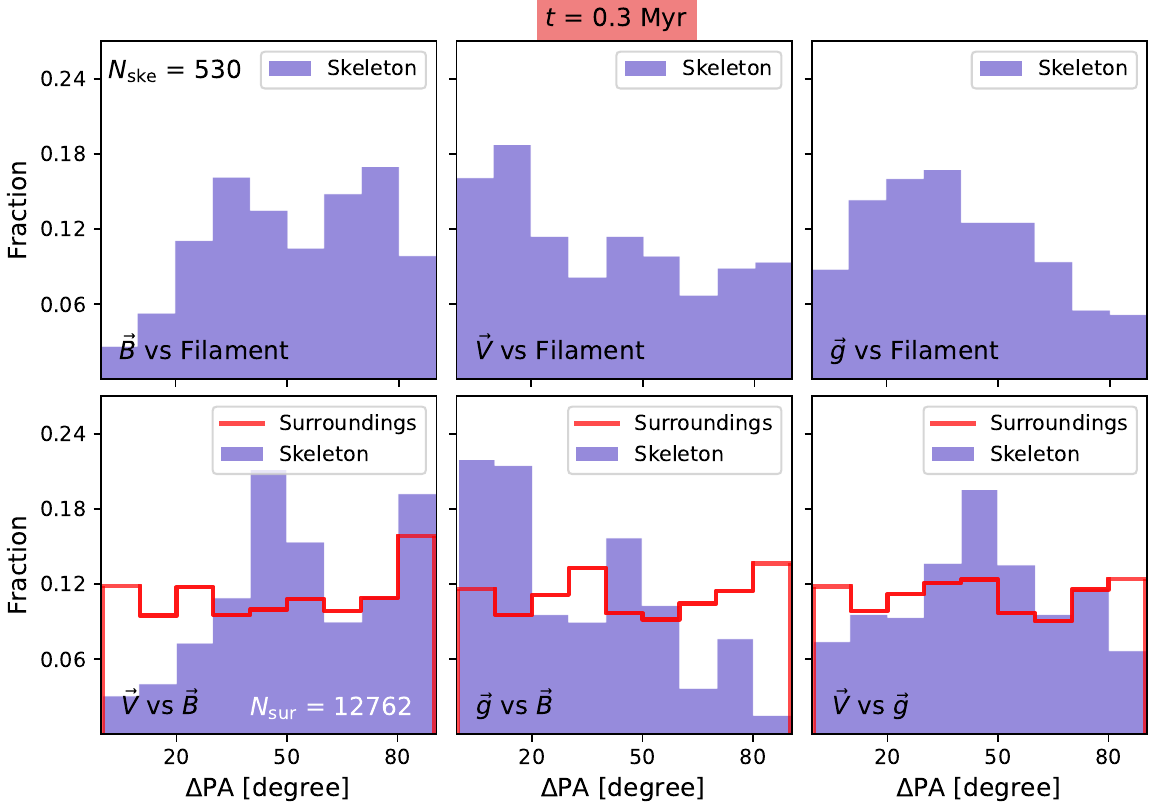}
 \includegraphics[angle=0,width= 0.7\textwidth,trim={0.0cm 0.0cm 0.0cm 0.0cm},clip]{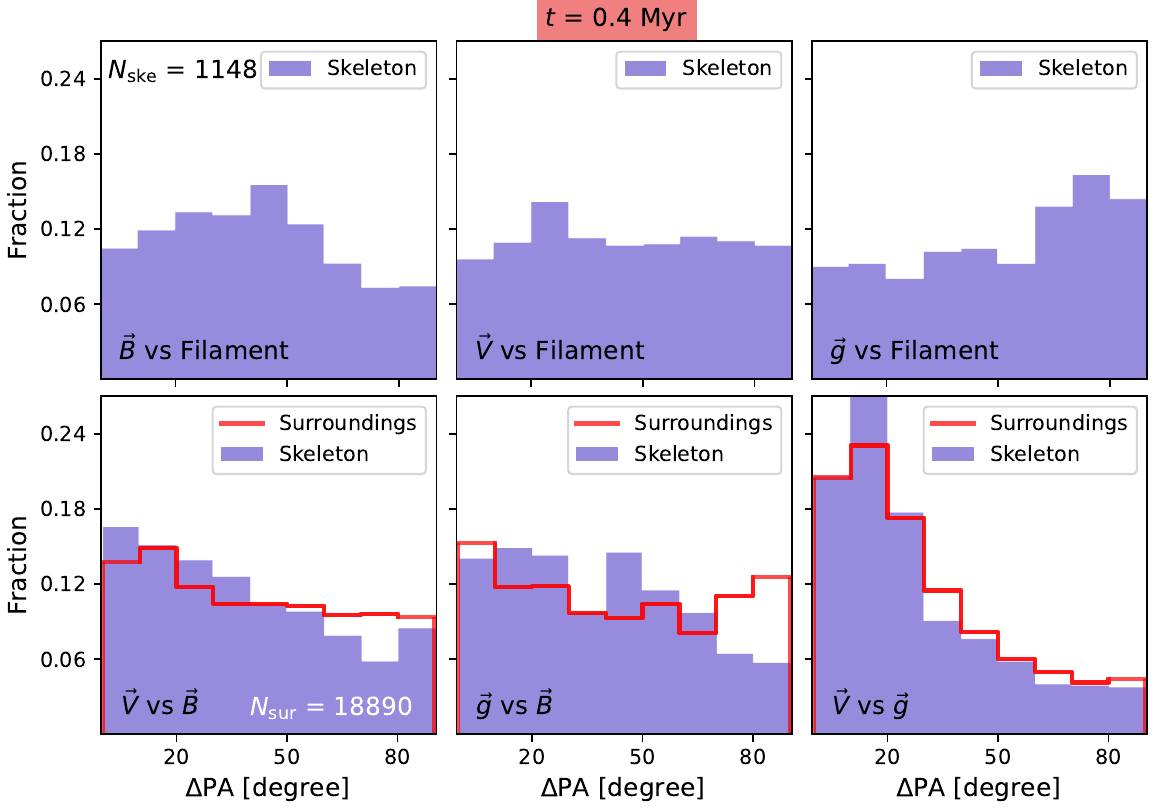}
\caption{Continued.}
\label{Delta_PA_dist_app}
\end{center}
\end{figure*}
\addtocounter{figure}{-1}
\begin{figure*}
\begin{center}
 \includegraphics[angle=0,width= 0.7\textwidth,trim={0.0cm 0.0cm 0.0cm 0.0cm},clip]{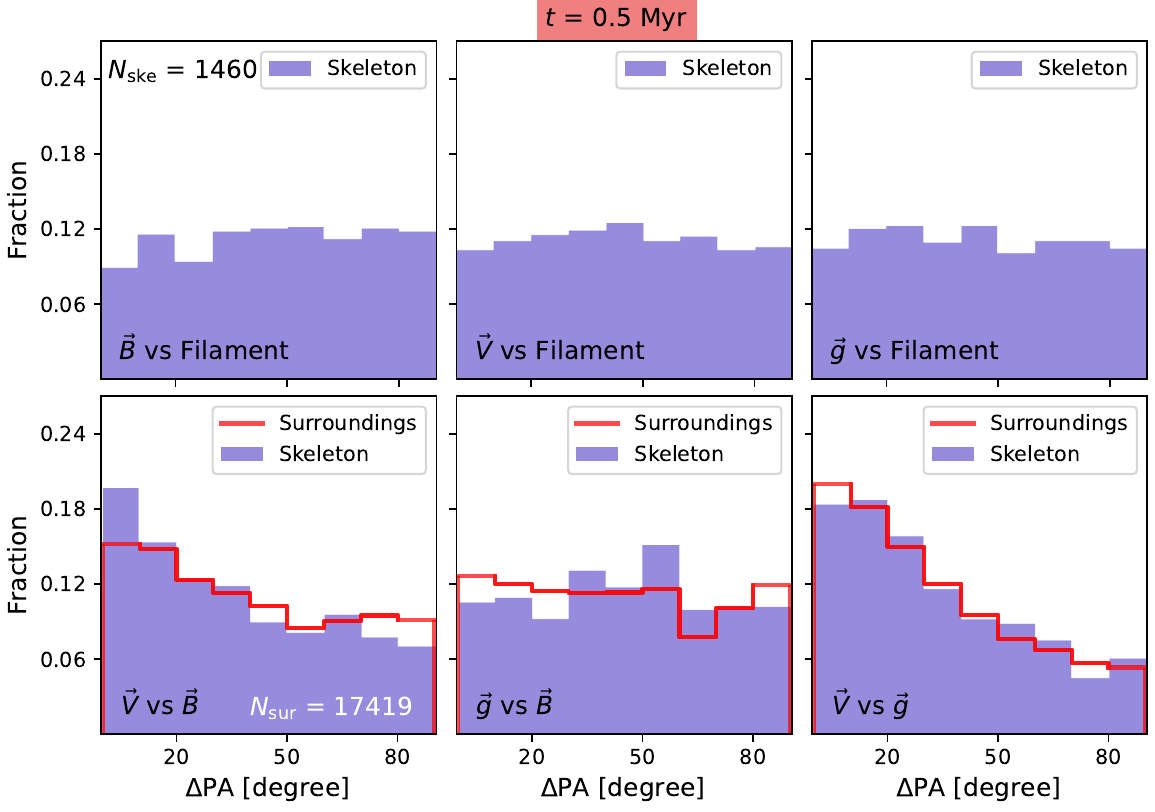}
 \includegraphics[angle=0,width= 0.7\textwidth,trim={0.0cm 0.0cm 0.0cm 0.0cm},clip]{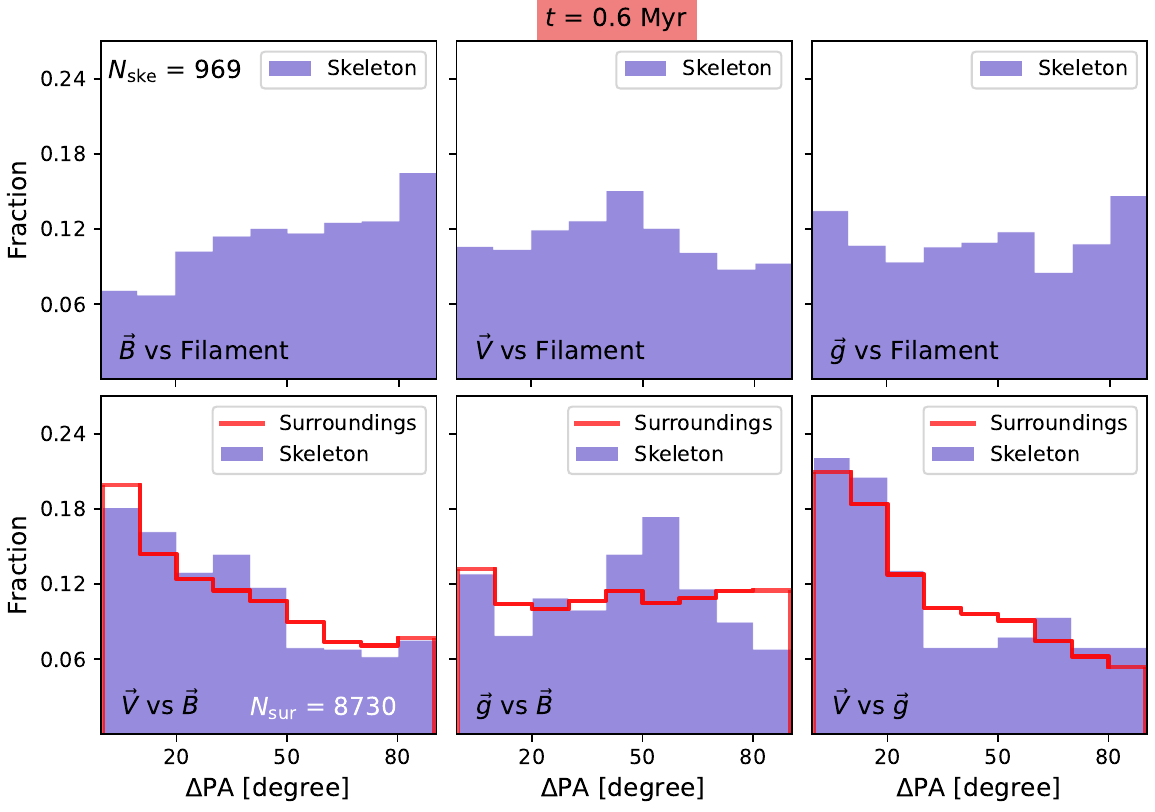}
\caption{Same as Figure~\ref{Delta_PA_dist}, but for $t$ = 0.3 to 0.6 Myr, respectively.}
\label{Delta_PA_dist_app}
\end{center}
\end{figure*}

\end{document}